%% file: main.tex
\documentclass[12pt]{article}
\usepackage[margin=1in]{geometry}
\usepackage[round,authoryear]{natbib}
\usepackage{amsmath}
\usepackage{amssymb}
\usepackage{graphicx}
\usepackage[labelfont=bf]{caption}
\usepackage{booktabs}
\usepackage{hyperref}
\usepackage{microtype}
\usepackage{authblk}  %
\usepackage{multirow} %
\usepackage{nameref} %
\usepackage{pdflscape}  %

\graphicspath{{./}}

\newcommand{\DIfull}{standard disruption index}  %

\newcommand{\DIshort}{\ensuremath{DI}}  %
\newcommand{\Ni}{N_F} %
\newcommand{\Nj}{N_B} %
\newcommand{\Nk}{N_R} %
\newcommand{\OneD}{\ensuremath{1D}}  %
\newcommand{\OneDfull}{One-Reference Disruption Index}  %

\title{Large scientific teams are more, not less, disruptive}

\date{}

\author[1]{Junming Huang}
\author[1,2]{Yu Xie}
\affil[1]{Paul and Marcia Wythes Center on Contemporary China, Princeton University}
\affil[2]{Department of Sociology, Princeton University}

\begin{document}

\maketitle

\input{numbers}
\input{introduction}

\input{results}

\input{discussion}

\input{data-methods}
\input{misc}

\bibliographystyle{abbrvnat}
\bibliography{references}

\clearpage

\input{supplementary}

\end{document}

%% file: numbers.tex
\newcommand{\OAsize}{47,134,871}
\newcommand{\ReflenFiveLoss}{11.44\%}
\newcommand{\ReflenTenLoss}{25.38\%}
\newcommand{\DIAucLow}{0.26}
\newcommand{\DIAucHigh}{0.79}
\newcommand{\DIAucTuring}{0.79}
\newcommand{\DIAucNobel}{0.73}
\newcommand{\DIAucLasker}{0.60}
\newcommand{\DIAucAps}{0.40}
\newcommand{\DIAucSboy}{0.26}
\newcommand{\DIAucTot}{0.55}
\newcommand{\OneDAucLow}{0.94}
\newcommand{\OneDAucHigh}{0.98}
\newcommand{\OneDAucTuring}{0.94}
\newcommand{\OneDAucNobel}{0.98}
\newcommand{\OneDAucLasker}{0.98}
\newcommand{\OneDAucAps}{0.98}
\newcommand{\OneDAucSboy}{0.95}
\newcommand{\OneDAucTot}{0.94}
\newcommand{\OneDSingleAuthor}{0.19}
\newcommand{\OneDMaxAuthors}{0.25}

%% file: introduction.tex
\begin{abstract}
Modern science has moved decisively toward larger and more collaborative research teams.
Yet an influential 2019 study reported that small teams are more disruptive than large teams, a conclusion difficult to reconcile with both the well-known trend toward increasing collaboration and the disproportionately large teams behind widely recognized disruptive research, such as work that has been awarded Nobel Prizes.
Here we show that the reported disruptive advantage of small teams is largely an artifact of measurement rather than a fact about teams. %
Analyzing about 60 million scientific publications, we find that larger teams cite more references and that the \DIfull{} declines mechanically as reference lists lengthen.
Equalizing reference-count distributions across team sizes reverses the negative gradient within the index's own framework. %
Evaluating each paper against one reference at a time yields a reference-robust measure that distinguishes recognized breakthroughs from ordinary papers more sharply and more consistently than the \DIfull{} across six independently curated benchmarks, including the Nobel Prize.
Under this validated measurement, scientific disruption increases with team size across four scientific domains and in every decade from the 1970s to the 2010s.
For scientific publications, we conclude that larger teams are more, not less, disruptive.  %
\end{abstract}

\section*{Introduction}
One distinct feature of modern science is its progression toward larger-scale and more collaborative research~\citep{1986-Price}.
Solo authorship has given way to multi-authored papers, average team size has risen steadily across fields, and collaboration now routinely crosses institutional, national, and disciplinary boundaries~\citep{2007-wuchty, 2009-Porter, 2012-Gazni, 2021-jones, 2024-Adams, 2025-james}.
Such collaboration brings together specialized knowledge across multiple domains that is difficult for any individual to command~\citep{2009-jones, 2021-jones}. 
At first glance, there appears to be a scientific payoff to collaboration, as larger teams tend to receive more citations and are disproportionately represented among highly visible papers
~\citep{2007-wuchty, 2014-lariviere, 2015-Abramo, 2025-james}. %
However, whether larger teams produce more disruptive science has been called into question by a highly influential paper published in \textit{Nature} in 2019, with the assertion that ``[l]arge teams develop and small teams disrupt science and technology''~\citep{2019-wu}. %
The negative association between team size and measured disruption (Figure~\ref{fig:overview}c) has been reported in other corpora~\citep{2023-WangChengJun} and has become one of the most widely discussed and highly cited findings in research on team science, contributing to broader concerns about an apparent decline in disruptive science and innovation~\citep{2023-park}. %
If the verdict is correct that large teams are indeed less disruptive than small teams, it is difficult to reconcile modern science's decisive shift toward larger and increasingly cross-boundary collaborations, the very organizational form least capable of transforming it. %

This assertion is at odds with the fact that, in recent decades, large teams have been behind many recognized scientific breakthroughs. %
Nobel Prize–winning papers are disproportionately produced by larger teams, a pattern documented in earlier research~\citep{2020-wei} and also evident in our analysis (Figure~\ref{fig:overview}a).  %
In this paper, we show that, for scientific publications, the reported disruptive advantage of small teams does not hold: it is largely an artifact of measurement rather than a fact about teams. Once systematic differences in citation practices are accounted for, that assertion does not survive. %

\section*{The Small-Team Advantage as a Methodological Artifact}
\begin{figure}[p]
    \centering
    \includegraphics[width=\textwidth]{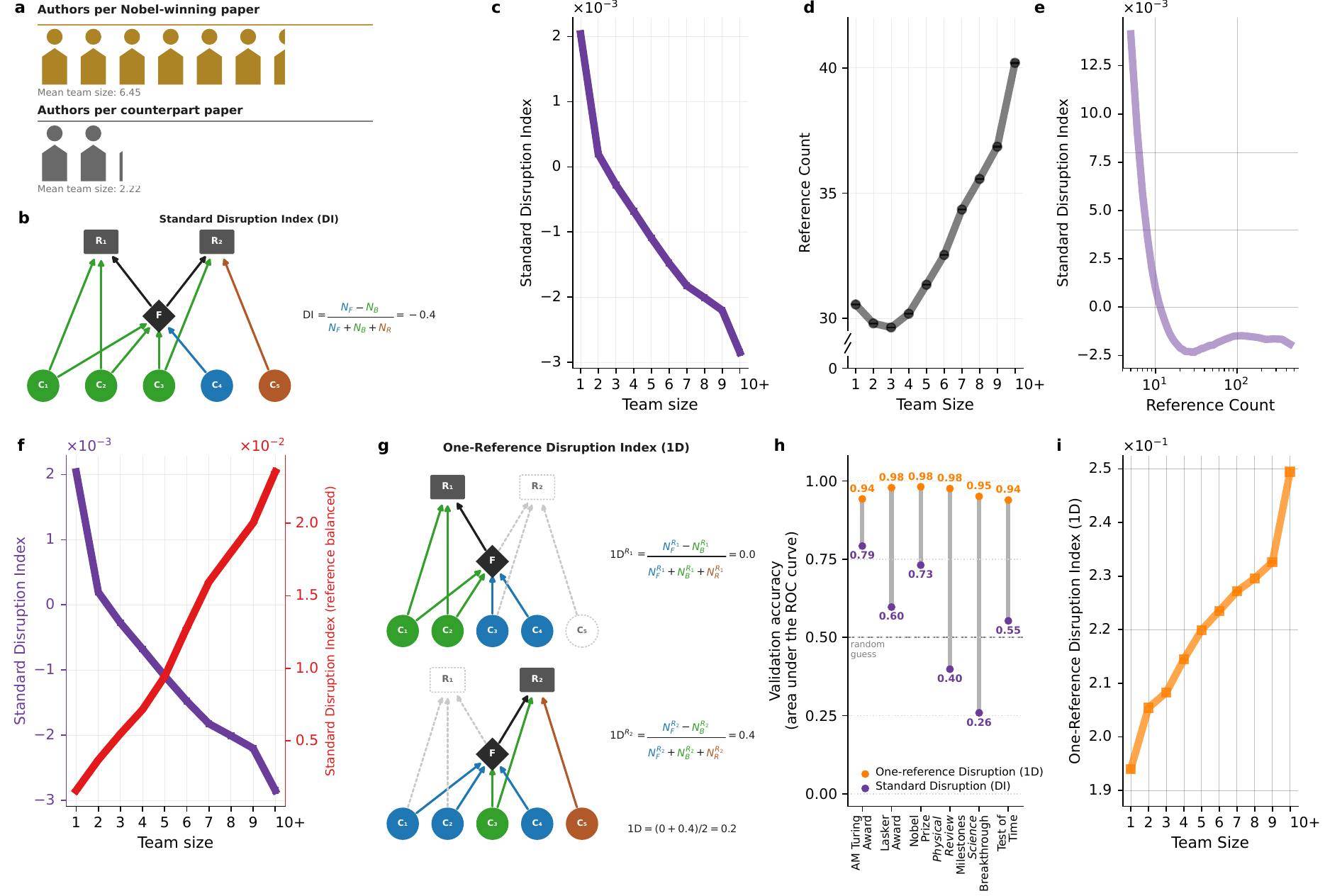}
    \caption{\textbf{Reassessing the apparent disruptive advantage of small scientific teams.}
    \textbf{(a)}, Nobel Prize-winning papers have larger teams than counterpart papers matched by domain and publication decade and weighted to the composition of the prize-winning set. Prize-winning papers have $6.45$ authors on average, against $2.22$ on average for counterparts.
    \textbf{(b)}, Standard disruption index with an example citation network.
    \textbf{(c)}, Mean \DIfull{} declines with team size. 
    \textbf{(d)}, Larger teams generally cite more references.
    \textbf{(e)}, Mean \DIfull{} varies with reference count, falling sharply and then flattening near zero for long lists.
    \textbf{(f)}, Mean \DIfull{} by team size before (purple, left axis) and after equalizing the reference-count distributions across team-size groups (red, right axis). The two axes are on different scales and the levels are not comparable; the comparison concerns the gradient direction, which reverses from negative to positive.
    \textbf{(g)}, The \OneDfull{} (\OneD{}) evaluates a focal paper against one reference at a time and averages the resulting scores, so the measure carries no cumulative penalty for citing more.
    \textbf{(h)}, Discrimination (area under the receiver operating characteristic curve) of the \DIfull{} (purple) and \OneD{} (orange) against six independently curated sets of recognized breakthroughs. \OneD{} more consistently distinguishes recognized breakthroughs from background papers. The dashed line marks chance-level discrimination.
    \textbf{(i)}, Under \OneD{}, disruption increases with team size. %
    Publication and citation data are from OpenAlex. %
    The baseline \DIshort{} scores (purple) are distributed with SciSciNet V2. %
    The reference-balanced series (red) is computed with our own implementation of the \DIfull{} (Data and Methods).
    }
    \label{fig:overview}
\end{figure}

The small-team advantage reported earlier by \citet{2019-wu} is largely a methodological artifact arising from a citation pattern that varies with team size. 
Larger teams cite more references (Figure~\ref{fig:overview}d), consistent with prior reports~\citep{2025-james}, and a paper's reference count strongly affects the \DIfull{} introduced by \citet{2017-funk} and subsequently applied at scale by \citet{2019-wu}.  %
The index is based on an intuitive distinction between displacing and extending prior knowledge~\citep{kuhn1962structure}: a paper is scored as disruptive when later work cites it while bypassing the studies it built on, and as developmental when later work cites it together with them.  %
However, the index is mechanically sensitive to the focal paper's reference count. The more references a paper cites, the more likely it is to include relevant prior studies, and the more likely subsequent papers are to cite the focal paper together with those predecessors, lowering its disruptive score.
A longer reference list also enlarges the pool of papers that engage the predecessors while bypassing the focal paper, compressing the index toward zero. %
These citation patterns mechanically drive the index downward and then cause it to level off near zero as reference lists lengthen (Figure~\ref{fig:overview}e), a methodological bias already documented in earlier research~\citep{2019-WuYan, 2020-Bornmann, 2021-ruan, 2024-leibel, 2024-petersen, 2024-yang}, with consequences for comparisons across team sizes~\citep{2025-petersen}.
The index therefore first penalizes the papers of larger, more heavily referencing teams with lower disruption scores and then, as reference lists lengthen further, becomes progressively less informative for distinguishing disruptive from developmental work. %
The same pattern arises even in randomized citation networks that retain no disruptive signal (Supplementary Information~\ref{si:mechanical}).  %
Yet the basic team-size question remains open: once we account for that methodological bias in the \DIfull{}, what relationship emerges between disruptiveness and team size? 

We answer this question by re-examining the relationship between team size and disruption using approximately 60 million scientific publications in OpenAlex.
We first demonstrate that the earlier reported negative association between team size and disruptiveness, which we replicate, is to a substantial degree mechanically driven, and becomes positive when we equalize reference count distributions across team-size groups inside the \DIfull{} framework (Figure~\ref{fig:overview}f).
Then, to measure disruption without the mechanical penalty for longer reference lists, we evaluate each paper's disruptiveness with respect to one reference at a time and average the resulting scores (\OneDfull{}, or simply \OneD{}, Figure~\ref{fig:overview}g).
We formulate uniform and citation-weighted versions of \OneD{}.
We then test 1D against six independently curated collections of recognized breakthroughs, on which it separates recognized from ordinary work more sharply and consistently than the \DIfull{} (Figure~\ref{fig:overview}h).
Using this validated measure, \OneD{}, we find that scientific disruption increases with team size (Figure~\ref{fig:overview}i), and this pattern holds across all four major scientific domains, in every decade from the 1970s to the 2010s, and irrespective of the choice of citation window (Supplementary Information~\ref{si:oneD-window}).

%% file: results.tex
\section*{Team Size and Scientific Disruption Revisited}
\begin{figure}[t]
    \centering
    \includegraphics[width=\textwidth]{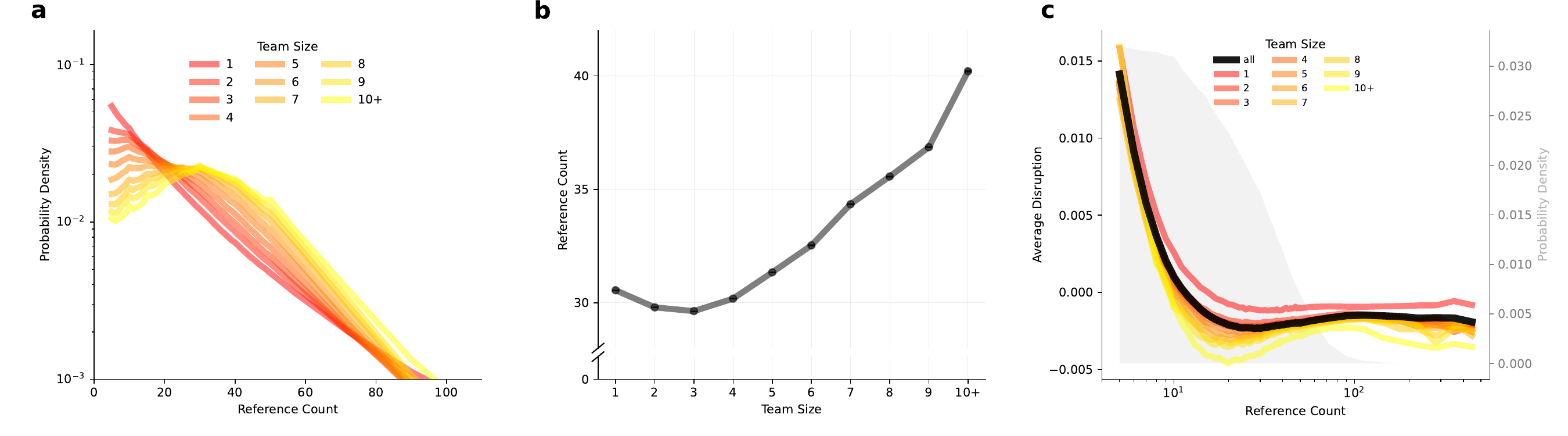}
    \caption{\textbf{Larger teams cite more references, while reference count accounts for far more of the \DIfull{} variation than team size does.}
    (a) Conditional distribution of reference count by team size groups. Single-author papers are concentrated at very low reference counts, while larger teams shift toward longer reference lists.
    (b) Mean reference count as a function of team size, showing a broad positive association across the bulk of the range.
    (c) Mean \DIshort{} as a function of reference count, overall (black) and separately by team size (colored). Across team-size groups, \DIshort{} shows a similar nonlinear decline as reference count increases: it drops sharply until reaching a minimum around reference count $\approx 30$, then plateaus near zero for higher reference counts. The remaining differences between groups are much smaller than the within-group variation across reference counts. The impact of reference count on \DIshort{} far exceeds that of team size.
    }
    \label{fig:distributions}
\end{figure}

\paragraph{Understanding the \DIfull{}.}
We analyze \OAsize{} scientific publications indexed in OpenAlex~\citep{2022-openalex} that list at least one author, cite at least five indexed references, and have been cited at least once~\citep{2024-holst, 2025-petersen} (see \nameref{sec:data-methods}). %
The paper-level \DIfull{} (\DIshort{}) was introduced by \citet{2017-funk} and applied at scale by \citet{2019-wu}. For a focal paper $F$,
\begin{equation}
    \DIshort{} = \frac{\Ni - \Nj}{\Ni + \Nj + \Nk},
    \label{eq:D}
\end{equation}
where $\Ni$ counts papers citing $F$ but none of its references, $\Nj$ counts papers citing both $F$ and at least one of its references, and $\Nk$ counts papers citing at least one reference of $F$ but not $F$ itself (Figure~\ref{fig:overview}b). \DIshort{} ranges from \(-1\) to \(+1\), with higher values indicating greater disruption~\citep{2017-funk, 2019-wu}. 
In practice, we use the precomputed \DIshort{} scores distributed with SciSciNet V2~\citep{lin2023sciscinet} as our baseline throughout, except where the index must be decomposed or recomputed (see \nameref{sec:data-methods}). Figure~\ref{fig:overview}c shows that those scores reproduce the negative association between team size and \DIshort{} reported by \citet{2019-wu}.
We proceed in three stages. We first show that the negative association between team size and \DIshort{} is generated by the index's structural dependence on reference count rather than a genuine difference in scientific disruption. We then propose and independently validate a reference-robust measure. Finally, using this measure, we show that the apparent disruptive advantage of small teams disappears and reverses.
We first argue that reference list length accounts for the observed negative relationship between team size and \DIshort{}. Two empirical observations inform this premise: larger teams systematically cite more references, and \DIshort{} declines as reference count grows~\citep{2025-petersen, 2024-petersen, 2024-leibel}.
As team size increases, the distribution of papers' reference counts shifts systematically toward longer reference lists (Figure~\ref{fig:distributions}a), producing a positive association between team size and mean reference count (Figure~\ref{fig:distributions}b) and, in turn, an artifactual negative association between team size and disruption as measured by \DIshort{}.

Because both team sizes and reference counts have increased over time~\citep{2007-wuchty, 2025-james}, their common temporal trend may produce an unconditional association between them.
We therefore examine the association conditional on publication decade. 
Regressing mean reference count in decade-by-team-size cells with decade fixed effects, we obtain a significantly positive slope along the plotted team-size scale ($0.50 \pm 0.09$ decade-size weighted, $0.33 \pm 0.12$ paper-level unweighted, see Supplementary Information~\ref{si:teamsize-reference}).

In turn, as reference count increases, the \DIfull{} falls steeply from its maximum at very low reference counts to a minimum around reference count $\approx 30$, and then plateaus near zero for higher reference counts, averaged over all papers (Figure~\ref{fig:distributions}c). 
Strikingly, the curves by different team sizes follow a very similar shape in Figure~\ref{fig:distributions}c.
The same result holds true across publication decades (Figure~\ref{fig:reflen-D-by-decade}).
In sum, given the positive association between team size and reference count (Figure~\ref{fig:distributions}b) and the negative association between reference count and disruption score (Figure~\ref{fig:distributions}c), it is plausible that much of the observed negative association between team size and \DIshort{} is accounted for by reference count.
\paragraph*{A mechanical decomposition shows how reference count biases \DIshort{}.}

\begin{figure}[t]
    \centering
    \includegraphics[width=\textwidth]{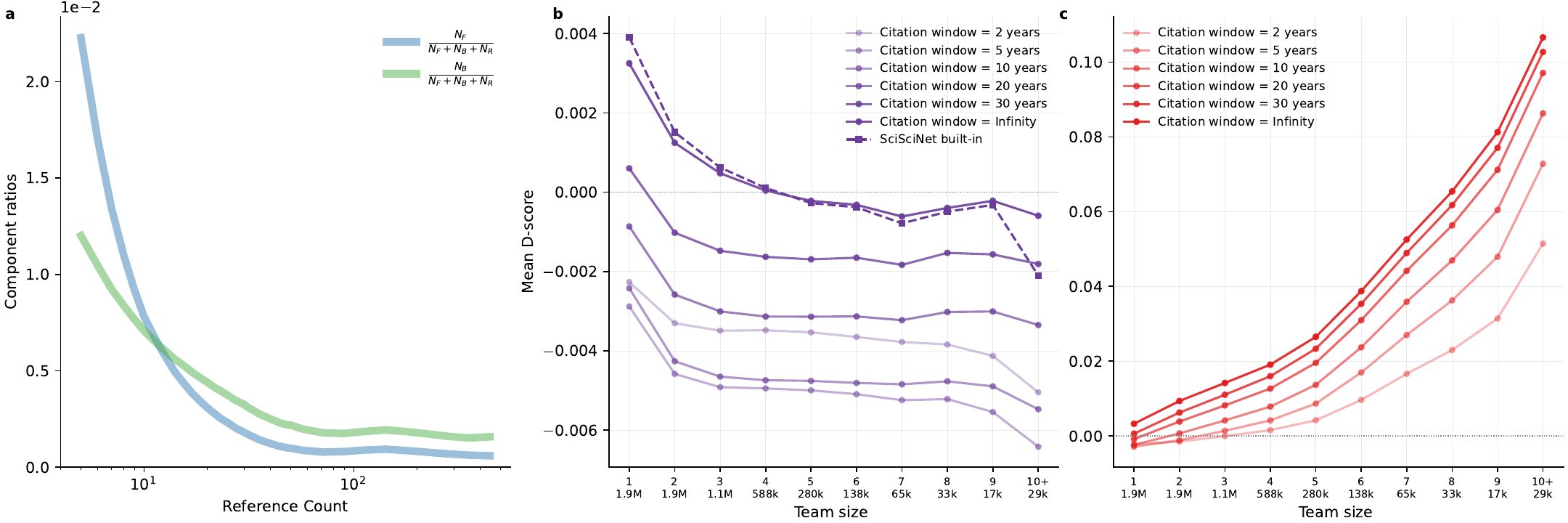}
    \caption{\textbf{Reference-count dependence of the \DIfull{}, and the team-size gradient before and after reference balancing.}
    \textbf{(a)} Mean normalized contributions of \(\Ni\) (blue) and \(\Nj\) (green) to \DIshort{} as functions of the focal paper's reference count. Both shares \(\Ni/(\Ni+\Nj+\Nk)\) and \(\Nj/(\Ni+\Nj+\Nk)\) decline as reference count increases, but \(\Ni/(\Ni+\Nj+\Nk)\) declines more rapidly. This difference in the rate of decline produces the decline in \(\DIshort{}=\Ni/(\Ni+\Nj+\Nk)-\Nj/(\Ni+\Nj+\Nk)\). The two shares cross at approximately 11 references, where \DIshort{} changes sign. Raw component counts and the corresponding decomposition in degree-preserving randomized citation networks are reported in Figure~\ref{fig:mechanism}.
    \textbf{(b)} Mean \DIshort{} as a function of team size in the original data, computed at citation windows of $2$, $5$, $10$, $20$, and $30$ years and without a window limit ($\infty$). The dashed line is the built-in \DIshort{} scores in SciSciNet, which mostly agrees with our implementation without a citation window limit. The gradient is negative at every citation window. 
    \textbf{(c)} Mean \DIshort{} after reference balancing. The gradient is positive at every citation window.
    The sample is the cumulative set of papers published before 1990, chosen because it can be observed at all window lengths. Team size is binned $1,\ldots,9,10+$, with bin sizes given below the labels. We use our own implementation of Equation~\eqref{eq:D}. 
    Because deleting references mechanically raises \DIshort{}, absolute score levels are not comparable between panels. The diagnostic comparison concerns the direction of the team-size gradient.
    }
    \label{fig:debiased_result_r6}
\end{figure}

This association between reference count and \DIshort{} arises mechanically from the construction of the index.
Recall that the \DIfull{} is defined as \(\frac{\Ni}{\Ni+\Nj+\Nk} - \frac{\Nj}{\Ni+\Nj+\Nk} \), where \(\Ni\) 
is the number of papers citing a focal paper alone, \(\Nj\) is the number of papers citing the focal paper together with its references, and \(\Nk\) is the number of papers citing only its references.  
Both normalized components decline as the focal paper's reference count increases, reflecting the dominance of \(\Nk\) in their common denominator (Figure~\ref{fig:debiased_result_r6}a).
Two consequences follow.
First, the normalized component of \(\Nj\) decreases more slowly than that of \(\Ni\), because a longer reference list increases the chance that a paper citing the focal paper also cites its references and is therefore counted in \(\Nj\) rather than \(\Ni\) (Figure~\ref{fig:debiased_result_r6}a). 
This differential decline pushes \DIshort{} downward, and the crossing of those two components at $11$ references marks where mean \DIshort{} turns negative (Figure~\ref{fig:distributions}c).
Second, growth in \(\Nk\) can cause the denominator to outpace the magnitude of the numerator, mechanically compressing \DIshort{} toward zero. This weaker pathway contributes at high reference counts and accounts for the plateauing tail in Figure~\ref{fig:distributions}c. This effect is mechanical because it requires no substantive differences in disruption, shown with randomized citation networks that we return in the Discussion.
This analysis does not mean that all empirical variation in \DIshort{} is mechanically induced. It shows, however, that differences in reference-count distributions create a substantial mechanical artifact in comparisons of \DIshort{} across groups, such as teams of different sizes. When reference counts differ systematically across groups, differences in \DIshort{} cannot be interpreted as valid differences in disruption.

\paragraph{Reference balancing overturns the team-size gradient for scientific publications at every citation window.}
Given the positive association between reference count and team size (Figure~\ref{fig:distributions}a), we now evaluate the behavior of \DIshort{} after equalizing reference-count distributions across team sizes. To this end, we construct a modified dataset in which the distribution of reference counts is forced to be the same across team-size groups.

Specifically, we take single-author papers (team size $=1$) as the baseline, because their cumulative distribution function (CDF) of reference count lies above those of larger teams over the main support (Figure~\ref{fig:reflen_cdf_by_teamsize}), so alignment can be achieved by deletion alone. %
Then, for each team size greater than one, we randomly remove references from focal papers until the reference-count distribution for that team size matches the baseline.
After this adjustment, the reference-count distributions are closely matched across team-size groups (Figure~\ref{fig:reflen_cdf_by_teamsize}b), removing most of the association between
team size and reference count (Supplementary Information~\ref{si:reference-balancing}).
We further recompute the \DIfull{} on the modified dataset, using citation windows of 2, 5, 10, 20, and 30 years, as well as an unrestricted window ($\infty$), where a $w$-year citation window counts only follow-up papers published within $w$ years after the focal paper (see \nameref{sec:data-methods}). %

The result is shown in Figure~\ref{fig:debiased_result_r6}b and Figure~\ref{fig:debiased_result_r6}c. In the original data, the \DIfull{} declines with team size at every citation window examined (Figure~\ref{fig:debiased_result_r6}b). 
The choice of citation window does not qualitatively change the declining trend, despite affecting the level of the curves. After reference balancing, the gradient is consistently positive at every window (Figure~\ref{fig:debiased_result_r6}c): disruption rises with team size across all windows. For this analysis, we limit to papers published through the 1980s because this sample can be observed at all windows. 
Analyses of nested samples extending through the 1990s, 2000s, and 2010s, using the corresponding observable windows, yield qualitatively similar results (Figure~\ref{fig:debiased_result_window}).

The reversal from a small-team advantage to a large-team advantage shows that, within the \DIfull{} framework, balancing reference counts is sufficient to eliminate the methodological artifact discussed earlier and reveal the advantage of large teams in scientific disruption.
This is a counterfactual demonstration rather than a causal identification: deleting references also alters the composition of the reference set and the counts \(\Nj\) and \(\Nk\), so no single pathway is isolated. We therefore do not interpret this exercise as identifying a causal effect of reference count.

\paragraph{\OneDfull{} (\OneD{}): a reference-robust paper-level measure.}
Next, we move from reference balancing, a group-level diagnostic design, to \OneDfull{} (\OneD{}), a paper-level disruption measure that by construction carries no cumulative penalty for citing more references (Figure~\ref{fig:overview}g). 
\OneD{} is defined as a revised version of Equation~\eqref{eq:D}, evaluating a focal paper against one reference at a time and averages the resulting scores. We consider two forms, uniform and citation-weighted, which differ in how references contribute to this average. Specifically, for a focal paper $F$ with reference list $R_F$, we first define its disruption relative to a reference $r$ as

\begin{equation}
  1D^{(r)} = \frac{\Ni^{(r)} - \Nj^{(r)}}{\Ni^{(r)} + \Nj^{(r)} + \Nk^{(r)}},
  \label{eq:oneD_r}
\end{equation}
where $\Ni^{(r)}$ counts papers citing $F$ but not $r$, $\Nj^{(r)}$ counts papers citing both $F$ and $r$, and $\Nk^{(r)}$ counts papers citing $r$ but not $F$.
The uniform \OneD{} gives each such reference equal weight:
\begin{equation}
  1D_{\mathrm{u}} = \frac{1}{|R_F|}\sum_{r \in R_F} 1D^{(r)}.
  \label{eq:oneD}
\end{equation}
The weighted \OneD{} instead weights each reference by its in-degree in the full citation network:
\begin{equation}
  1D_{\mathrm{w}} = \frac{\sum_{r \in R_F} c_r 1D^{(r)}}{\sum_{r \in R_F} c_r},
  \label{eq:oneD_weighted}
\end{equation}
where $c_r$ is the number of papers in the full citation network that cite $r$, including $F$, regardless of citation window. Both forms are probability expectations of $1D^{(r)}$ in Equation~\eqref{eq:oneD_r}: the uniform \OneD{} selects $r$ with equal probability $1/|R_F|$, whereas the weighted \OneD{} selects $r$ with probability $c_r/\sum_{s\in R_F}c_s$. Either form can be computed by exact enumeration or approximated by averaging independent draws from its corresponding reference distribution. If $R_F$ is empty, the score is defined as $\mathrm{NaN}$. In the empirical results below, \OneD{} denotes the uniform form.

For the uniform form, this expectation can be accurately approximated using Monte Carlo averaging on a small random sample of references. 
We run $m=30$ independent draws with replacement, with each run randomly drawing one reference from $R_F$, and then compute a Monte Carlo estimator
\begin{equation}
  \widehat{\OneD}_{\mathrm{MC}} = \frac{1}{m}\sum_{b=1}^{m}\OneD{}^{(r_b)}.
  \label{eq:oneD_approx}
\end{equation}
The sampling error of this Monte Carlo estimator declines with $m$ as \(\sigma_F / \sqrt{m}\), where \(\sigma_F\) is the standard deviation of \(1D^{r}\) across the references of $F$. At the group level this error is negligible (three orders of magnitude smaller, see SI~\ref{si:oneD-convergence}), and therefore we use Equation~\eqref{eq:oneD_approx} for large-scale group comparisons like team size analysis for efficiency. %

Because each reference-level score involves exactly one reference, both mechanical channels identified above are blocked by construction: neither \(\Nk^{(r)}\) nor the transfer between \(\Ni^{(r)}\) and \(\Nj^{(r)}\) grows as the focal paper's reference list lengthens. What this removes is the cumulative penalty, not every association between \OneD{} and reference count; mean \OneD{} still varies with reference count in our data (Supplementary Information~\ref{si:oneD-robustness}).

\paragraph{Independent empirical validity using expert-curated benchmarks.}
\begin{figure}[!t]
    \centering
    \includegraphics[width=\textwidth]{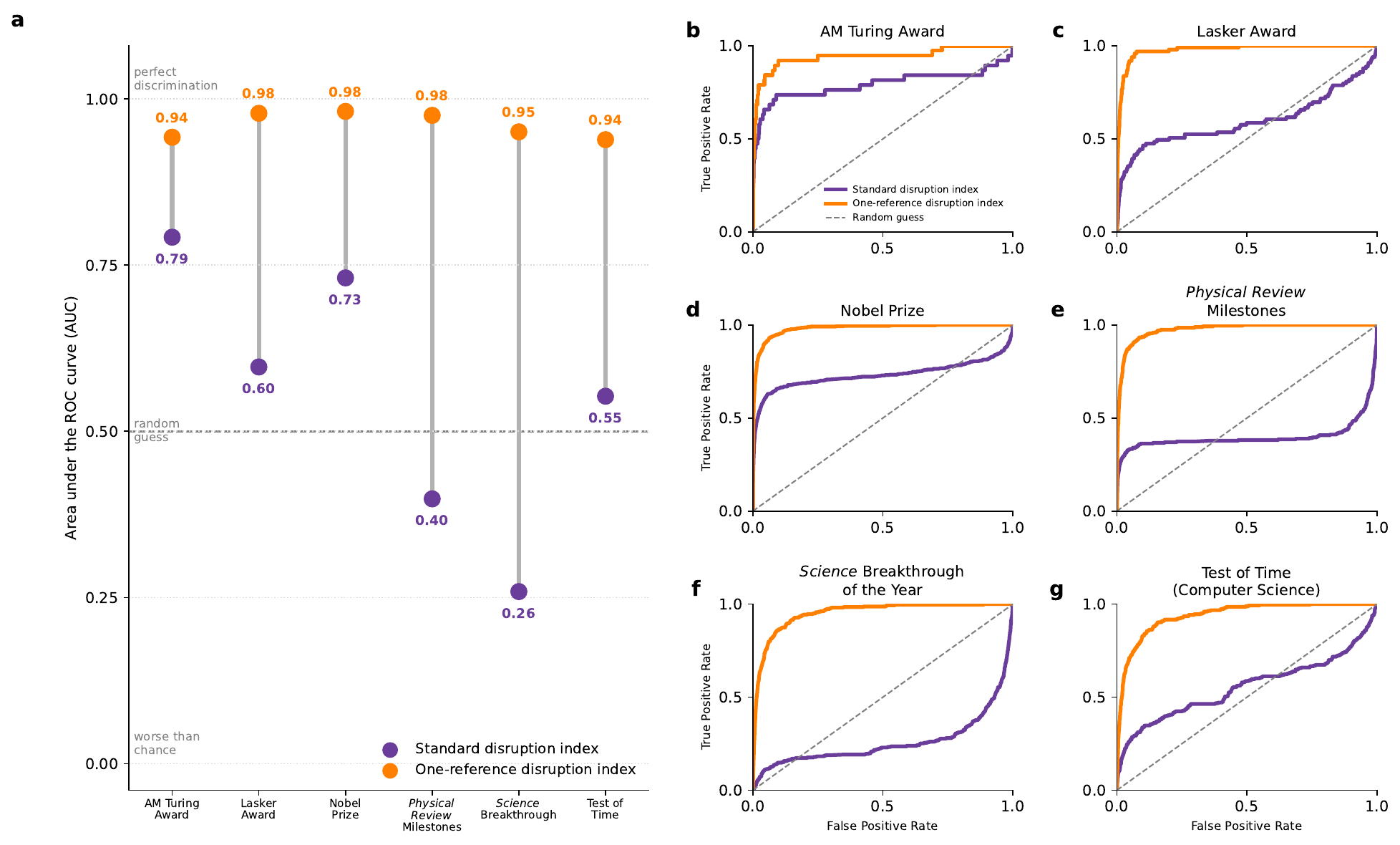}
    \caption{
    \textbf{Independent benchmarks of expert-recognized breakthroughs.}
    The \DIfull{} (SciSciNet built-in scores, purple) and \OneD{} (orange), both without a citation-window limit, are tested against six benchmarks on a binary task separating positives (papers in a curated set) from negatives (all other scored papers, excluding the union of the six benchmarks).
    \textbf{(a)} Area under the Receiver Operating Characteristic curve (AUC) for each benchmark. Horizontal segments connect the AUC values of the two measures within each benchmark. The dashed line marks the random ranking (\(\mathrm{AUC}=0.5\)). \OneD{} shows consistently high discrimination across benchmarks (AUC \(=\OneDAucLow{}\)--\(\OneDAucHigh{}\)), whereas the \DIfull{} is substantially weaker and more variable (AUC \(=\DIAucLow{}\)--\(\DIAucHigh{}\)), falling below \(0.5\) for \textit{Physical Review} Milestones and the \textit{Science} Breakthrough of the Year. AUC below \(0.5\) signals that \DIshort{} scores rank recognized papers below background papers more often than above them within these benchmark-specific samples.
    \textbf{(b--g)} The corresponding Receiver Operating Characteristic (ROC) curves for the ACM A.M. Turing Award, Lasker Award, Nobel Prize, \textit{Physical Review} Milestones, \textit{Science} Breakthrough of the Year, and Test of Time awards. The dashed diagonal indicates chance-level discrimination.}
    \label{fig:ground_truth_roc}
\end{figure}

We assess the empirical validity of \OneD{} by testing whether it is better than the \DIfull{} in distinguishing work that expert communities have independently recognized as major scientific breakthroughs.
We evaluate the \DIfull{} and \OneD{} against six independent expert-curated benchmarks spanning multiple scientific domains: the ACM A.M.\ Turing Award (computer science), the Nobel Prize (physics, chemistry, and physiology or medicine), the Lasker Awards (Biomedicine), Test of Time awards (computer science), \textit{Physical Review} Milestone papers (physics), and the \textit{Science} Breakthrough of the Year~\citep{2018-li, ioannidis2020work, 2025-yy, 2023-WangShiyun} (see \nameref{sec:data-methods}).
For each benchmark, we compare \DIfull{} scores distributed with SciSciNet and \OneD{} scores computed using an unrestricted citation window, treating papers included in the benchmark as positives and all other scored papers outside all benchmarks as negatives. %
Because the positives and negatives are extremely imbalanced, with tens to hundreds of recognized papers against a background of tens of millions (Table \ref{tab:benchmarks}), we summarize discrimination performance by the receiver operating characteristic (ROC) curve and its area (AUC), which remains interpretable under the extreme class imbalance by evaluating relative ranking rather than classification accuracy at a fixed threshold.

\OneD{} consistently discriminates expert-recognized breakthroughs from the background literature across all six benchmarks, with AUC values between \(\OneDAucLow\) and \(\OneDAucHigh\) (Figure~\ref{fig:ground_truth_roc}).
In contrast, the \DIfull{} is substantially weaker and less consistent. It shows moderate discrimination for the Turing Award (AUC \(=\DIAucTuring\)) and the Nobel Prize (AUC \(=\DIAucNobel\)), only slight discrimination for the Lasker Awards (AUC \(=\DIAucLasker\)), near-chance discrimination for Test of Time awards (AUC \(=\DIAucTot\)), and AUC values below \(0.5\) for \textit{Physical Review} Milestones (AUC \(=\DIAucAps\)) and \textit{Science} Breakthrough of the Year (AUC \(=\DIAucSboy\)). On these last two benchmarks, the \DIfull{} more often ranks a randomly selected recognized paper below a randomly selected background paper than above it. 
Across all these six heterogeneous benchmarks, \OneD{} provides substantially stronger and more consistent discrimination than the \DIfull{}. 
This convergence across independently curated sets lends credence to the greater validity and robustness of \OneD{} relative to the \DIfull{} in ranking recognized work.%

\paragraph{Large scientific teams are more disruptive}

\begin{figure}[!t]
    \centering
    \includegraphics[width=\textwidth]{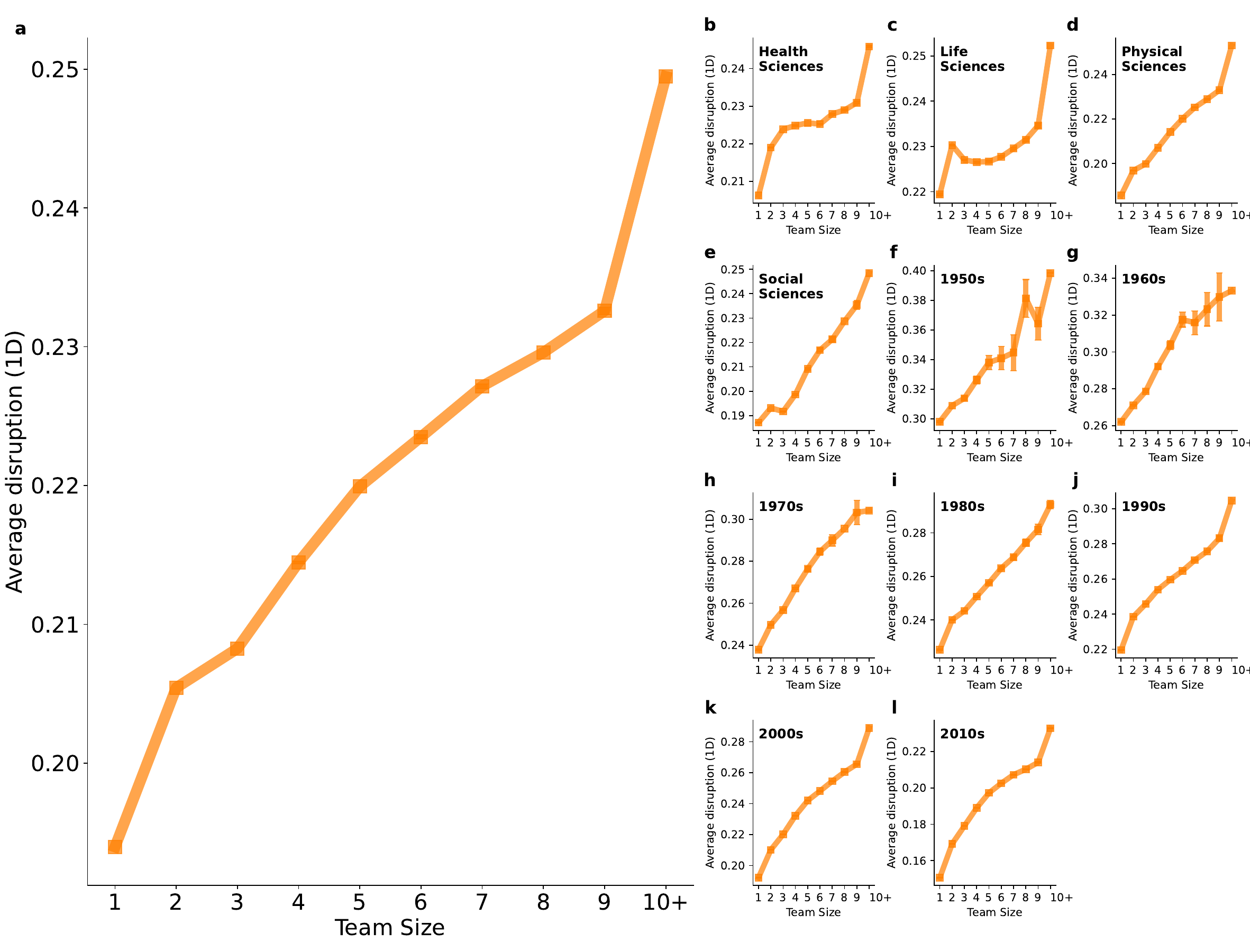}
    \caption{
    \textbf{After accounting for reference count bias, scientific disruption increases with team size.}
    \textbf{(a)} Mean \OneD{} as a function of team size. Disruption rises across the range and is highest for the largest teams. The standard errors are narrower than the plotted curves because of the large sample size and are therefore not visible.
    \textbf{(b--e)} Mean \OneD{} increases with team size in each of four major scientific domains, namely Health Sciences, Life Sciences, Physical Sciences, and Social Sciences.
    \textbf{(f--k)} Mean \OneD{} increases with team size in each decade from 1950s to 2010s. 
    }
    \label{fig:debiased_result}
\end{figure}

Having established the validity of \OneD{} independently of team size, we use it to characterize how disruption varies with team size. Under \OneD{}, mean disruption clearly increases with team size, rising from \(\OneDSingleAuthor\) for single-author papers and reaching its highest value of \(\OneDMaxAuthors\) for the largest team size bin of ten or more (Figure~\ref{fig:debiased_result}a).
The pattern is not confined to a particular field or period. It holds across all four major scientific domains (Figure~\ref{fig:debiased_result}b--e) and in every decade from the 1950s to the 2010s (Figure~\ref{fig:debiased_result}f--k), and at every citation window choice (Figure~\ref{fig:oneD_teamsize_windows}). 
This pattern is the inversion of the well-cited negative gradient, which our preceding analyses show to be highly sensitive to the mechanical coupling between team size and reference count, by construction of \DIshort{}.
Under a measure that passes independent validity tests, scientific disruption in publications increases, rather than decreases, with team size.
These findings have important implications both for research on team science and for the broader literature that relies on the \DIfull{} as a measure of scientific disruption.

%% file: discussion.tex
\section*{Discussion and Conclusion}
We show that the widely accepted negative association between team size and disruption~\citep{2019-wu} is, to a large extent, mechanically driven. 
Larger teams systematically produce papers with longer reference lists, a relationship that persists within publication decades and therefore cannot be attributed solely to parallel temporal growth. 
We demonstrate that reference count accounts for far more of the \DIfull{} variation than team size does, and this dependence is largely structural rather than substantive, arising from the construction of the \DIfull{} itself. 
Reference balancing provides direct evidence for this artifact: the negative team-size gradient disappears and becomes positive when the reference-count distributions of different team-size groups are brought into close alignment. 
Then we introduce a reference-robust paper-level measure, \OneDfull{} (\OneD{}), that removes the mechanical penalty created when additional references accumulate in the paper-level denominator. 
Validation of uniform \OneD{} against six external benchmarks reveals a concern with the \DIfull{}: it fails to consistently rank recognized papers above background papers, whereas \OneD{} does so across all six benchmarks.
Under uniform \OneD{}, scientific disruption increases with team size across major scientific domains and publication decades. 
Taken together, these findings suggest that the canonical claim that large teams are less disruptive than small teams is largely a measurement artifact. 
Once the mechanical dependence on reference count is addressed, larger scientific teams are more, not less, disruptive.

Our results help explain a longstanding puzzle in the disruption literature: 
Scientific work has shifted decisively toward larger and increasingly cross-boundary collaborations, yet the \DIfull{} has portrayed large teams as primarily developmental rather than disruptive.
Many Nobel Prize-winning papers, despite being widely recognized as major breakthroughs, do not receive substantially higher \DIfull{} scores than ordinary papers~\citep{2020-wei}. Because prize-winning papers tend to come from larger teams, which cite more references, the \DIfull{} assigns many of them scores that are difficult to reconcile with their recognized status. The same holds across the other benchmarks: on two of them, the \DIfull{} ranks a recognized paper below a background paper more often than above it. 
Under \OneD{}, recognized papers are much more clearly distinguished from background papers than under the \DIfull{}, and their disruption scores are correspondingly more consistent with their status as recognized scientific advances.
This positive association between team size and scientific disruption, revealed by \OneD{}, aligns more closely with a broader empirical literature showing that larger teams often possess characteristics conducive to transformative science, including coping with the burden of knowledge~\citep{2009-jones}, novel combinations of ideas~\citep{2021-jones}, and diverse backgrounds~\citep{2025-yang}.

Our results should be interpreted cautiously.
We do not treat citation impact as equivalent to disruption, but expert recognition provides an external empirical criterion for assessing whether a disruption measure ranks consequential work in a consistent direction. \OneD{} discriminates consistently across all six benchmarks while reference count alone does not (Supplementary Information~\ref{si:refcount-baseline}), which indicates that \OneD{} is not simply a restatement of reference count. \OneD{} nevertheless remains empirically associated with reference count: it removes the mechanical penalty rather than every relationship.
We also do not interpret the positive gradient causally. Team size may proxy division of labor, knowledge integration, access to specialized resources, or other features of collective research~\citep{2021-zeng}. Our design does not identify which mechanism produces the observed pattern or estimate the effect of adding an author to a paper.
Reference balancing is a diagnostic intervention on the observed citation network, not a causal design. Deleting references changes both the number and identities of a paper's recorded predecessors, and the remaining association between team size and disruption may reflect topic choice, resources, selection into collaboration, or other unobserved differences.
Data quality is an especially important limitation. Missing or incomplete records in bibliographic databases can alter citation-network statistics and may do so systematically rather than randomly~\citep{2024-holst}, which is not unique to OpenAlex.
More broadly, our findings show that measurement design can change central conclusions about the organization of science. Before disruption indicators are used to evaluate fields, institutions, or collaboration policies, their mechanical dependencies should be examined and their behavior validated against empirical cases external to the metric itself. For scientific publications, the evidence points to greater, not lesser, disruption among papers produced by larger teams.

%% file: data-methods.tex
\section*{Data and Methods}
\label{sec:data-methods}

\paragraph{Data and analysis population.}
We used the OpenAlex 2025 snapshot, which links publications, authors, citations, and research topics across scientific fields~\citep{2022-openalex}. For each work, we extracted its publication year, author count (team size), references, incoming citations, and primary topic. 
We restrict the analysis to \OAsize{} focal papers published before 2020, with at least one author, at least five references, and at least one citation recorded in the snapshot. We impose no lower cutoff on publication year but exclude papers published in or after 2020 because they have not yet accumulated sufficient citations. We put no filter on citing and cited papers. We operationalize a paper's ``team size'' as its number of listed authors, following the common practice without suggesting that this is the only definition of collaboration~\citep{1997-katz}. Reference count is the number of listed references.
For analyses by field, we assigned each paper to the single top-level domain of its OpenAlex \texttt{primary\_topic}.%

We exclude focal papers with fewer than five indexed references, because very short reference lists are disproportionately associated with editorials, comments, and other non-research items rather than substantive research articles~\citep{2025-petersen}.
Selecting this cutoff requires balancing data quality against broad coverage. A low threshold admits many non-research works~\citep{2025-petersen} and can produce spuriously high \DIshort{} scores if zero-reference records are included~\citep{2024-holst}. A high threshold, however, would discard many substantive research articles. For example, the ten-reference threshold suggested by \citet{2025-petersen} would exclude \ReflenTenLoss{} of eligible focal papers. We therefore choose five-reference threshold as a compromise between those two, which excludes \ReflenFiveLoss{} of eligible focal papers (Figure~\ref{fig:reflen-distri}). %

\paragraph{Relation to alternative disruption measures.}
The disruption literature has proposed several responses to the \DIfull{}, including the negative association between team size and disruption largely disappearing once reference count and citation count are taken into account~\citep{2025-petersen}. These concerns have motivated a series of revisions to the \DIfull{}, including downweighting highly cited predecessors or requiring stronger co-citation evidence~\citep{2020-Bornmann,2023-deng}, removing or rescaling the $\Nk$ component~\citep{2019-WuYan,2024-leibel,2024-yang}, and requiring disruptive works to remain directly cited~\citep{2025-deng, 2026-deng}.
A distinct approach learns separate past- and future-context vectors for each paper from the citation network and defines disruptiveness as their distance~\citep{2025-yy}.
Recent research has shown that disruption scores are sensitive to reference count and citation inflation and that this sensitivity can alter temporal and team-size comparisons~\citep{2021-ruan,2024-petersen,2025-petersen}.
Different from those approaches, \OneD{} prevents the number of references in a focal paper from mechanically expanding a single paper-level comparison set, while retaining the expected distinction among later papers that cite the focal work and/or its references.

\paragraph{Comparison of Nobel and counterpart papers.}
Figure~\ref{fig:overview}a compares the team sizes of Nobel Prize-winning papers with matched counterpart papers. %
We match each Nobel paper to counterpart papers with the same primary domain and publication decade, and then reweight the matched counterpart papers so that all Nobel papers and all counterpart papers are identically distributed over domains and publication decades. That yields a fair comparison on team size between the two groups.

\paragraph{Disruption measures and citation windows.}
For most analyses involving the \DIfull{}~\citep{2017-funk,2019-wu}, our default input is the precomputed \DIshort{} scores distributed with SciSciNet V2~\citep{lin2023sciscinet}, which links \DIshort{} scores to OpenAlex works as built-in values. SciSciNet V2's author team overlaps with that of ~\citet{2019-wu}, which ensures that the central comparison does not depend on choices in our own implementation. Whenever $\Ni+\Nj+\Nk=0$, we consider \DIshort{} in the indeterminate form $0/0$ and label such a paper with missing score.

The only exception is analysis that decomposes \DIshort{} or filters the citing papers. We use our own implementation of Equation~\eqref{eq:D} to examine the $\Ni$, $\Nj$, and $\Nk$ components (Figure~\ref{fig:debiased_result_r6}a), to recompute \DIshort{} after reference balancing (Figure~\ref{fig:debiased_result_r6}c), and to conduct supplementary robustness analyses in different citation windows (Figure~\ref{fig:debiased_result_r6}b,c).
A $w$-year citation window counts only follow-up papers published within $w$ years of the focal paper~\citep{2025-lin}. We calculate \DIshort{} using citation windows of 2, 5, 10, 20, and 30 years. Without such a finite citation window, our implementation of \DIshort{} closely agrees with the SciSciNet V2 builtin scores on the analyzed population (Figure~\ref{fig:debiased_result_r6}b).
The comparison in Figure~\ref{fig:debiased_result_r6}b,c uses a common pre-1990 cohort so that every citation window can be observed for the same publication cohort. Comparisons on other cohorts and corresponding citation windows are reported in Supplementary Information~\ref{si:reference-balancing}.

Every reported \OneD{} analysis uses its Monte Carlo estimator (Equation~\eqref{eq:oneD_approx}) with $m=30$ independent draws per focal paper, and no citation-window limit unless stated otherwise. %
Replications of Figure~\ref{fig:debiased_result} under finite citation windows yield consistent conclusions (Supplementary Information~\ref{si:oneD-window}).

\paragraph{Team size and reference count.}
We group papers into the team-size categories $1,2,\ldots,9$, and $10+$ authors.
Both team size and reference count increased historically, so their pooled association might partly reflect a common time trend~\citep{2007-wuchty,2025-james}.
We therefore summarize the association using decade-by-team-size cells.
For each cell, we calculate mean reference count and fit a regression with publication-decade fixed effects; the main specification weights cells by the number of papers they represent, and alternative weighting schemes are reported in the Supplementary Information.
This design compares the ordered team-size groups after absorbing differences in mean reference count across publication decades.
It does not remove within-decade historical change and is not a causal model.
Because the final group contains every paper with ten or more authors and is represented by its lower boundary, the coefficient is a descriptive slope along the team-size scale used in the figure, not the individual marginal effect of adding one author to a paper.
The full equation, cell means, weights, top-coding rule, and sensitivity analyses are provided in the Supplementary Information.

\paragraph{Reference-balancing demonstration.}
We use controlled deletion to test whether the negative team-size gradient in the \DIfull{} persists after the reference-count distributions of different team-size groups are brought into close alignment.
We first bin papers into reference count categories $1,2,3,\ldots,49,50+$, where the last bin pools all papers with 50 or more references.
Let $Pr(c \mid s)$ denote the probability of observing reference count $c$ among papers with team size $s$. We empirically observe its cumulative distribution function $F_s (C) = \sum_{c=1}^{C}\Pr(c\mid s)$ satisfies,

\[
\widehat F_s(C) \leq \widehat F_1(C), \forall C=1,2,3,\ldots,49,50+.
\]
That allows us to choose single-author papers as the target distribution of reference count, and align other team size groups to this target distribution with \textbf{deletion only}, rather than adding references that a paper did not cite. For every team size $s > 1$, we randomly pick such papers and randomly remove their references until $\widehat F_s(c) \approx \widehat F_1(c)$ for every reference count bin $c > 1$. We then recompute our implementation of Equation~\eqref{eq:D} on the resulting citation network, and report the average result over 12 random replicas.
The resulting comparison should be interpreted as a measurement diagnostic rather than causal identification. %
In the balancing figures, error bars summarize one standard error across the 12 replica means. The detailed description of deletion procedure, binned coupling, treatment of the upper tail, distributed sampling procedure, and balance diagnostics are described in the Supplementary Information.

\paragraph{Validation benchmarks.}
We validate the \DIfull{} and \OneD{} using six external collections of recognized research: ACM A.M.\ Turing Award papers, Lasker Award papers, Nobel Prize-winning papers, \textit{Physical Review} Milestone papers, \textit{Science} Breakthroughs of the Year, and Test of Time (computer science) award papers~\citep{2018-li,ioannidis2020work,2025-yy,2023-WangShiyun}.
Those source lists were deduplicated, adjudicated, and linked to OpenAlex papers before analysis. For consistency, we apply the same data filter of other analyses to remove 267 papers that are not in OpenAlex, 56 papers without a \DIshort{} score, 140 papers with fewer than five references, 1 paper without citations, and 6 papers published after 2020, ending up with 1,469 eligible recognized papers.
These filters use a fixed priority and are not marginal effects.
Table~\ref{tab:benchmarks} gives a brief description of all the benchmark datasets.

\begin{table}[ht]
    \centering
    \caption{\textbf{Validation with expert-curated breakthrough benchmarks.}}
    \label{tab:benchmarks}
    \begin{tabular}{llllll}
    \toprule
    Benchmark & Field & \shortstack[l]{Number\\of papers} & \shortstack[l]{Number of \\eligible papers} & \DIshort{} AUC & \OneD{} AUC\\
    \midrule
    \shortstack[l]{ACM A.M.\\ Turing Award}         & \shortstack[l]{Computer\\ science} & 79           & 38  & 0.792  & \textbf{0.942} \\
    \midrule
    Lasker Awards                & Biomedicine & 114          & 99  & 0.597     & \textbf{0.979}  \\
    \midrule
    Nobel Prize  & \shortstack[l]{Physics, chemistry,\\ and physiology\\ or medicine} & 778          & 468 & 0.731     & \textbf{0.981} \\
    \midrule
    \shortstack[l]{\textit{Physical Review}\\ Milestone} & Physics & 300          & 270 & 0.398     & \textbf{0.975} \\
    \midrule
    \shortstack[l]{\textit{Science} \\Breakthroughs \\of the Year} & Multidisciplinary & 334          & 318 & 0.259     & \textbf{0.951} \\
    \midrule
    Test of Time    & \shortstack[l]{Computer\\ science} & 334          & 276 & 0.553   & \textbf{0.939} \\   
    \bottomrule
    \multicolumn{2}{l}{Total} & 1,939 & 1,469 & 0.555 & \textbf{0.961} \\
    \bottomrule
    \end{tabular}
\end{table}

For each benchmark, built-in \DIshort{} and Monte Carlo \OneD{} were evaluated on exactly the same positive and negative paper groups. The positive group consists of recognized papers in this benchmark, and the negative group are all other papers in the analysis population excluding the union of all benchmark collections. We calculate Receiver Operating Characteristic curve for \DIshort{} and \OneD{} to measure their ability to distinguish positive and negative papers, and report the area under curve (AUC) as a scalar performance indicator.  %

The benchmark labels played no role in defining or tuning either measure. As a diagnostic, we evaluate reference count itself as a predictor on the same positive and negative groups. Reference count does not discriminate consistently: its AUC ranges from
0.31 to 0.65 across the six benchmarks, falling below 0.5 for the Nobel Prize and the Turing Award. Because a monotonically increasing transformation of reference count preserves these values and a decreasing one replaces each by approximately one minus its value, no monotonic transformation of reference count can reproduce the consistently high discrimination of \OneD{}. Therefore the consistently high discrimination of \OneD{} cannot be reduced to a repackaged reference-count predictor, even though long reference lists are more common among recognized papers in some collections (Supplementary Information~\ref{si:refcount-baseline}, Table~\ref{si:tab:benchmarks}).

\paragraph{Statistical reporting.}
Unless stated otherwise, line plots report arithmetic means for the displayed groups and error bars show one standard error, not bootstrap 95\% confidence intervals.

%% file: misc.tex
\section*{Data availability}

OpenAlex data are publicly available at \url{https://openalex.org}. SciSciNet V2 data are publicly available at \url{https://github.com/kellogg-cssi/SciSciNet}.

\section*{Code availability}

Analysis code will be deposited at GitHub. %

\section*{Acknowledgements}

The authors thank Lingfei Wu and Dashun Wang for insightful comments and for providing computational details of the \DIfull{} algorithm, and thank Alexander Chen (Princeton University) for helping collect the curated recognized paper lists.

\section*{Author contributions}

YX and JH conceptualized and designed the study, and developed the statistics and algorithms. JH implemented the methods. Both authors performed the analyses. JH created visualizations of the results and wrote the first draft. Both authors revised the manuscript.

\section*{Competing interests}

The authors declare no competing interests.

%% file: supplementary.tex
\appendix

\renewcommand{\thefigure}{S\arabic{figure}}
\renewcommand{\thetable}{S\arabic{table}}
\renewcommand{\theequation}{S\arabic{equation}}

\setcounter{figure}{0}
\setcounter{table}{0}
\setcounter{equation}{0}

\section*{Supplementary Information}

\section{Eligibility and coverage diagnostics}
\label{si:data-construction}

The focal-paper eligibility rules are specified in Methods. Here we quantify the selectivity of the requirement that a focal paper have at least one observed citation.
Table~\ref{tab:uncited-share-team-size} reports the fraction of otherwise linkable papers with no citation in the snapshot.
The excluded share is highest for single-author papers and declines with team size.
The main team-size curves should therefore be interpreted as relationships among papers that have entered the observed citation network, not among all publications.

\begin{table}[htbp]
\centering
\caption{Share of otherwise linkable papers with no recorded citation, by team size. The diagnostic population has joinable metadata, at least one author, at least one linked reference, and a valid publication year. The citation filter is not applied until after these shares are calculated.}
\label{tab:uncited-share-team-size}
\begin{tabular}{lrrr}
\toprule
Team size & Papers & Never cited & Share (\%) \\
\midrule
1   & 19,209,453 & 7,493,525 & 39.0 \\
2   & 17,712,094 & 4,375,596 & 24.7 \\
3   & 15,007,790 & 3,261,082 & 21.7 \\
4   & 11,061,447 & 2,164,902 & 19.6 \\
5   &  7,750,067 & 1,412,435 & 18.2 \\
6   &  5,501,941 &   950,735 & 17.3 \\
7   &  3,511,772 &   552,850 & 15.7 \\
8   &  2,394,805 &   366,926 & 15.3 \\
9   &  1,585,996 &   232,881 & 14.7 \\
$10+$ & 3,902,403 & 565,645 & 14.5 \\
\bottomrule
\end{tabular}
\end{table}

\section{Team size and reference count}
\label{si:teamsize-reference}

\begin{figure}[ht]
    \centering
    \includegraphics[width=\textwidth]{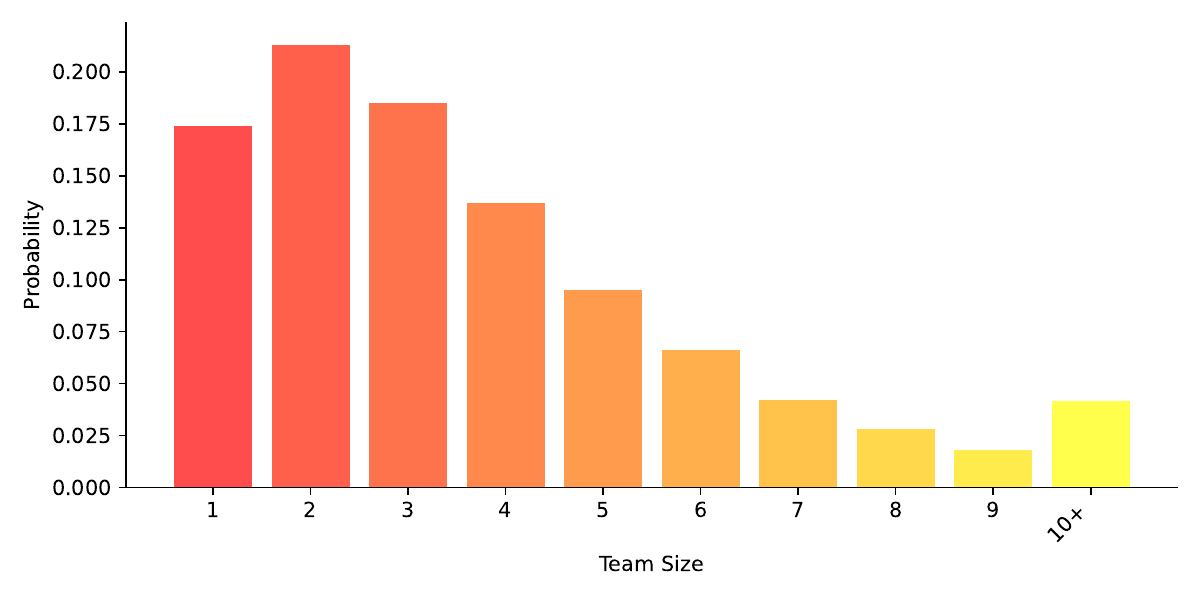}
    \caption{Distribution of team size in OpenAlex.}
    \label{fig:teamsize-distri}
\end{figure}

\begin{figure}[h]
    \centering
    \includegraphics[width=\textwidth]{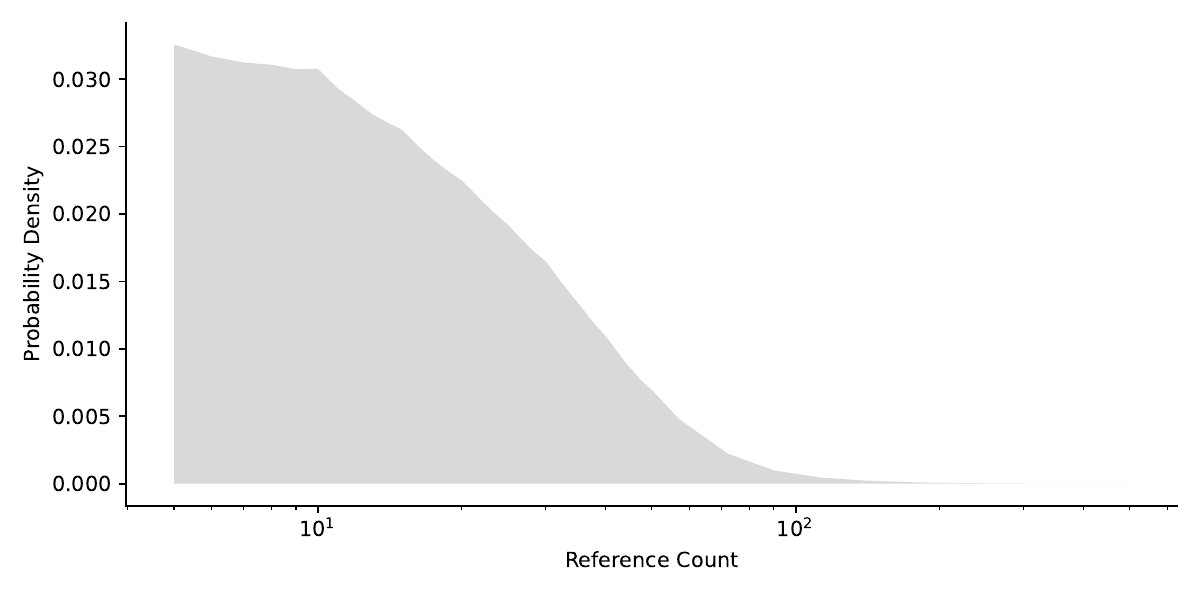}
    \caption{Distribution of reference count in OpenAlex.}
    \label{fig:reflen-distri}
\end{figure}

Let $\bar R_{dk}$ denote mean reference count in publication period $d$ and team-size bin $k$, and let $n_{dk}$ be the corresponding number of papers.
We estimate
\begin{equation}
  \bar R_{dk}=\alpha_d+\beta \cdot k+\varepsilon_{dk},
  \label{eq:si-reflen-team-fe}
\end{equation}
where $k=1,\ldots,9,10$, with the open-ended $10+$ group represented by its lower boundary.
The regression is fitted to $8 \times 10=80$ decade--team-size cells, covering 47,134,871 papers.
Papers published before 1950 are pooled into one ``decade'', followed by decades from the 1950s through the 2010s.

We report two weighting schemes.
The first gives every cell in decade $d$ weight $n_d=\sum_k n_{dk}$, so team-size bins receive equal weight within a decade and decades contribute in proportion to their numbers of papers.
The second gives each cell weight $n_{dk}$, which is equivalent to giving every paper equal weight.
Table~\ref{tab:reference-regression-weights} shows that the pooled descriptive slope is positive under both choices.

\begin{table}[tb]
\centering
\caption{Decade-fixed-effects regression of mean reference count on team size. Standard errors use the 80 cell means, with 71 residual degrees of freedom. The coefficient is a bin-level descriptive slope along the plotted team-size scale, not the causal effect of adding an author.}
\label{tab:reference-regression-weights}
\begin{tabular}{lrrr}
\toprule
Weighting & $\hat\beta$ & SE & Two-sided $p$ \\
\midrule
Decade-size weighted & 0.5235 & 0.0857 & $4.8\times10^{-8}$ \\
Paper-level unweighted & 0.3201 & 0.0936 & 0.0010 \\
\bottomrule
\end{tabular}
\end{table}

\begin{table}[tb]
\centering
\footnotesize
\setlength{\tabcolsep}{4pt}
\caption{Mean reference count by publication decade and team size. Together with Table~\ref{tab:teamsize_cells_n}, these are the cells of the fixed-effects regression in Equation~\eqref{eq:si-reflen-team-fe}. Column headings give the team size; the bin $10^{+}$ collects all teams of ten or more and is assigned $k=10$ in the regression.}
\label{tab:teamsize_cells_reflen}
\begin{tabular}{lrrrrrrrrrr}
\toprule
 \multirow{2}{*}{Decade} & \multicolumn{10}{c}{Team size} \\
\cmidrule(l){2-11}
 & 1 & 2 & 3 & 4 & 5 & 6 & 7 & 8 & 9 & $10^{+}$ \\
\midrule
Pre-1950 & 15.669 & 12.285 & 12.154 & 12.716 & 13.784 & 23.315 & 21.533 & 19.000 & 66.683 & 52.607 \\
1950s & 16.953 & 14.271 & 13.448 & 13.991 & 15.031 & 16.302 & 18.546 & 21.943 & 26.394 & 29.792 \\
1960s & 18.282 & 15.771 & 14.854 & 15.564 & 15.765 & 16.263 & 16.562 & 18.892 & 22.300 & 22.941 \\
1970s & 20.622 & 18.322 & 17.370 & 17.422 & 18.051 & 18.464 & 19.081 & 19.397 & 19.114 & 18.545 \\
1980s & 24.689 & 21.897 & 20.357 & 20.135 & 20.535 & 21.063 & 21.607 & 22.170 & 22.558 & 22.341 \\
1990s & 27.962 & 26.193 & 24.351 & 24.000 & 24.486 & 25.006 & 26.021 & 26.606 & 27.194 & 27.857 \\
2000s & 32.899 & 30.425 & 28.325 & 27.851 & 28.376 & 29.125 & 30.432 & 31.440 & 32.470 & 34.794 \\
2010s & 35.754 & 35.406 & 34.760 & 34.652 & 35.296 & 36.141 & 37.830 & 38.755 & 39.839 & 43.083 \\
\bottomrule
\end{tabular}
\end{table}

\begin{table}[tb]
\centering
\footnotesize
\setlength{\tabcolsep}{4pt}
\caption{Number of papers by publication decade and team size for the cells whose mean reference counts are given in Table~\ref{tab:teamsize_cells_reflen}. These counts permit construction of the decade-size and paper-level weights reported in Table~\ref{tab:reference-regression-weights}.}
\label{tab:teamsize_cells_n}
\begin{tabular}{lrrrrrrrrrr}
\toprule
 \multirow{2}{*}{Decade} & \multicolumn{10}{c}{Team size} \\
\cmidrule(l){2-11}
 & 1 & 2 & 3 & 4 & 5 & 6 & 7 & 8 & 9 & $10^{+}$ \\
\midrule
Pre-1950 & 132\,519 & 52\,918 & 14\,396 & 3\,411 & 927 & 346 & 92 & 51 & 41 & 84 \\
1950s & 129\,219 & 82\,163 & 34\,625 & 11\,407 & 3\,507 & 1\,234 & 454 & 229 & 94 & 173 \\
1960s & 300\,320 & 253\,021 & 119\,244 & 43\,122 & 13\,918 & 5\,088 & 1\,912 & 859 & 400 & 866 \\
1970s & 550\,648 & 580\,357 & 331\,581 & 152\,519 & 62\,018 & 25\,990 & 10\,633 & 4\,942 & 2\,497 & 5\,016 \\
1980s & 816\,994 & 923\,726 & 640\,593 & 377\,728 & 199\,383 & 104\,858 & 51\,767 & 26\,962 & 13\,614 & 23\,102 \\
1990s & 1\,138\,671 & 1\,388\,291 & 1\,087\,678 & 749\,876 & 481\,944 & 306\,570 & 175\,808 & 103\,441 & 58\,875 & 100\,831 \\
2000s & 2\,128\,248 & 2\,658\,457 & 2\,318\,621 & 1\,706\,765 & 1\,168\,966 & 813\,560 & 507\,977 & 328\,402 & 206\,320 & 420\,618 \\
2010s & 3\,003\,456 & 4\,106\,538 & 4\,163\,955 & 3\,407\,284 & 2\,538\,161 & 1\,869\,608 & 1\,237\,865 & 859\,688 & 576\,149 & 1\,410\,680 \\
\bottomrule
\end{tabular}
\end{table}

The pooled descriptive slope is positive under both weighting schemes, but does not imply that every decade-specific association is positive.

\section{Mechanical dependence of the \DIfull{} on reference count}
\label{si:mechanical}

\begin{figure}[h]
    \centering
    \includegraphics[width=\textwidth]{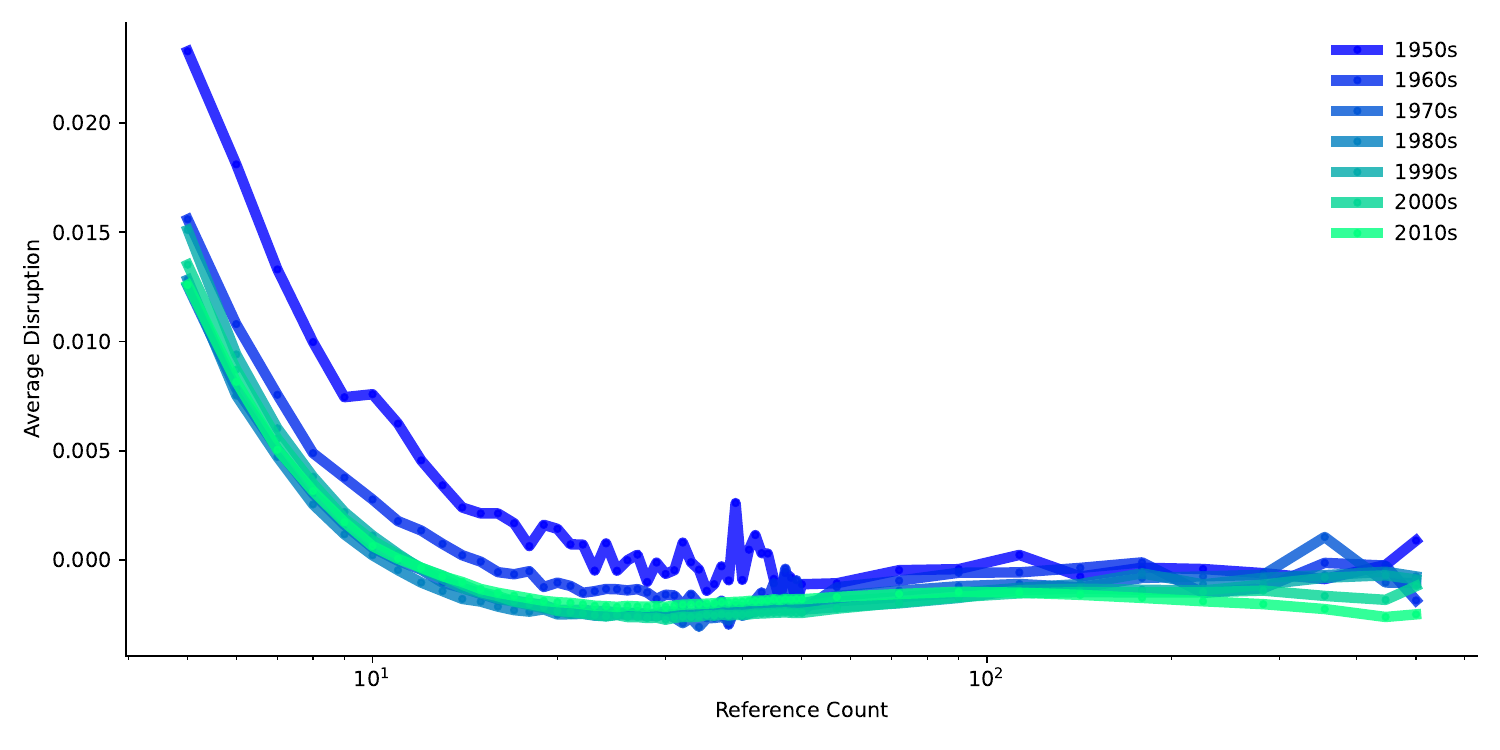}
    \caption{Mean \DIshort{} as a function of reference count, plotted separately by decade from the 1970s through the 2010s. The negative association is visible within publication decades and is therefore not explained solely by historical changes in publication and citation practices.}
    \label{fig:reflen-D-by-decade}
\end{figure}

Equation~\eqref{eq:D} can be written as the difference between two normalized components,
\begin{equation}
  \DIshort{}=\frac{\Ni}{\Ni+\Nj+\Nk}-\frac{\Nj}{\Ni+\Nj+\Nk}.
\end{equation}
Across focal papers, the denominator expands as reference count increases because more predecessors create more opportunities for later following papers to cite at least one item in the reference set.
A longer reference list also increases the opportunity for a later paper that cites the focal paper to cite one of its predecessors, placing that later paper in $\Nj$ rather than $\Ni$.
This is a distributional comparison across focal papers, without assuming that a longer reference list of one focal paper is a superset of the shorter list of another focal paper.

In the observed network, both normalized components decline with reference count, but $\Ni/(\Ni+\Nj+\Nk)$ declines more rapidly than $\Nj/(\Ni+\Nj+\Nk)$.
The two components cross between approximately 11 and 12 references, where mean \DIshort{} changes sign.
At high reference counts, the score approaches a floor near zero because the growth of $\Nk$ makes the denominator large relative to the bounded difference $\Ni-\Nj$.
The empirical curve therefore declines and then converges near zero.%

\begin{figure}[t]
    \centering
    \includegraphics[width=\textwidth]{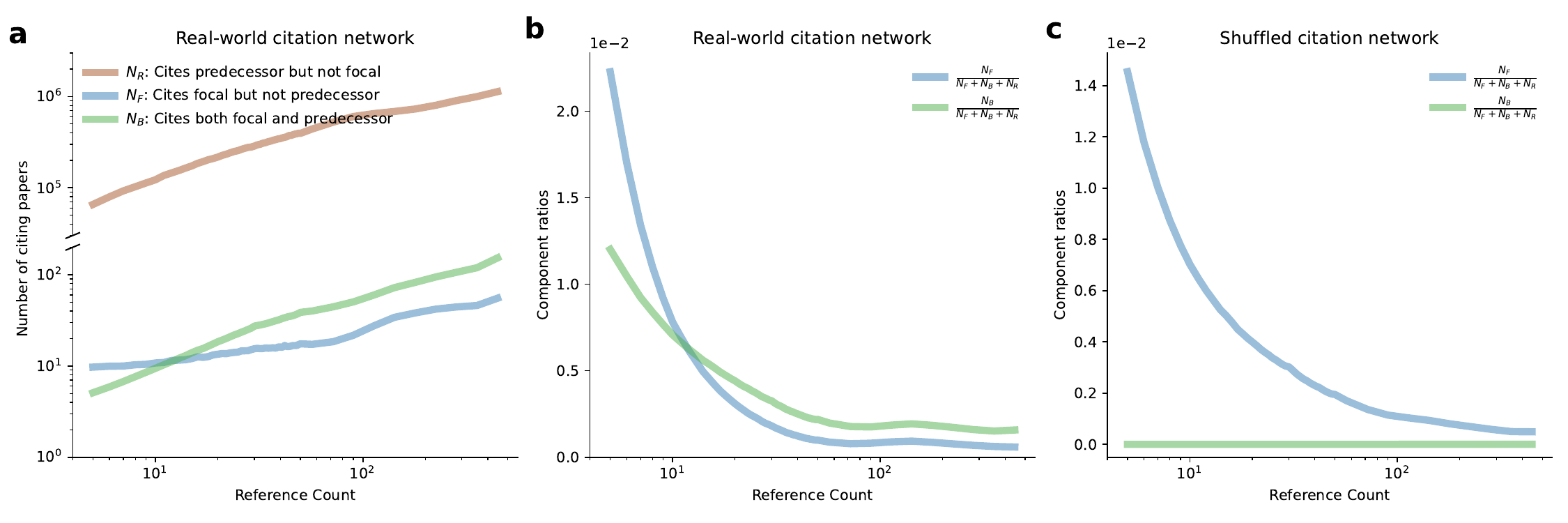}
    \caption{\textbf{The \DIfull{} is mechanically sensitive to reference count.}
    \textbf{(a)} Mean component counts $\Ni$, $\Nj$, and $\Nk$ as functions of focal-paper reference count in the observed citation network; $\Nk$ is much larger than the other components.
    \textbf{(b)} The normalized shares $\Ni/(\Ni+\Nj+\Nk)$ and $\Nj/(\Ni+\Nj+\Nk)$ in the observed network. Both decline with reference count, but the $\Ni$ share declines faster, producing the downward \DIshort{} curve and its sign change.
    \textbf{(c)} The corresponding shares in randomized citation networks. The $\Nj$ share remains near zero after co-citation structure is disrupted, whereas the $\Ni$ share still declines. Panels use our implementation of Equation~\eqref{eq:D} because the decomposition requires the component counts.}
    \label{fig:mechanism}
\end{figure}

\subsection{Degree-preserving randomized citation networks}
\label{si:null-construction}

We use randomized citation networks to test whether a declining \DIshort{}--reference-count curve can arise after any genuine disruption information is removed.
Specifically, we permute the ends of citing--cited links while retaining the topology of the citation graph.
The diagnostic uses stub permutation rather than double-edge swaps or unconstrained random rewiring for efficiency.
The observed sparse citation matrix is partitioned into \(100 \times 100 = 10{,}000\) shards, defined by 100 citing-community blocks and 100 cited-community blocks. Papers are assigned to communities based on the last two digits of their 64-bit paper IDs. At this large scale, this procedure yields an approximately random assignment, as we find no statistically significant association between these digits and any meaningful paper characteristics.
Within each shard, the row and column totals provide out-degree and in-degree targets.
Source and target stub lists are formed from the degree requirements, the target stubs are randomly permuted, and paired stubs propose new edges.
Duplicate paper pairs are coalesced and self-links are removed in diagonal shards.

Coalescing can leave residual degree requirements, particularly when a shard contains a very high-degree node.
The implementation therefore pre-allocates difficult high-degree endpoints and repeats the permute-and-coalesce step on the remaining residuals.
This construction preserves the community-block edge structure and targets the observed in-degree and out-degree sequence within every shard while changing citation partners.
It is not a global configuration-model draw.

A shard is accepted when it reaches its original edge count or differs from it by no more than ten (10) links, i.e., shuffling error lower than $0.005\%$ as each shard contains $\sim 2\times10^{5}$ citation links.
The resulting network is therefore degree-preserving to a small numerical tolerance rather than exactly degree-preserving for every node.
Against a huge network containing hundreds of millions of papers, the aggregate discrepancy is negligible, but we report it because it is part of the implemented null.
We generated ten parallel randomized networks.
Due to random fluctuations, a very small set of grouped cells is represented by nine rather than ten non-empty replicates; these cells account for less than 0.001\% of the weighted observations.

\begin{figure}[t]
    \centering
    \includegraphics[width=0.6\textwidth]{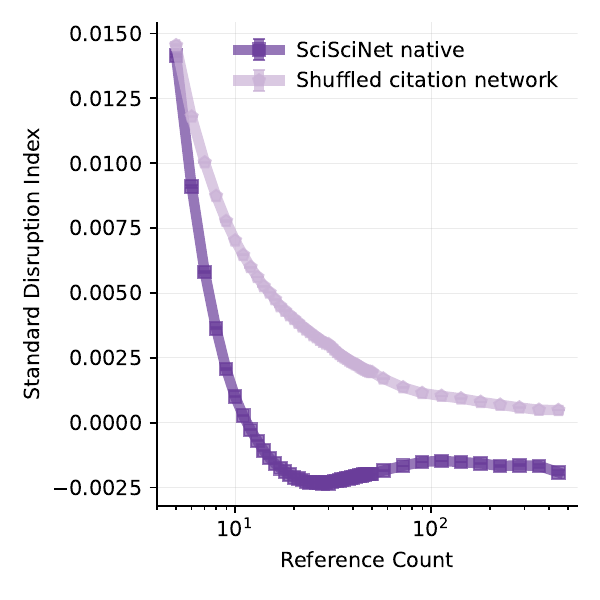}
    \caption{\textbf{The decline of \DIshort{} with reference count is reproduced in citation networks whose citation partners have been randomized.}
    The published SciSciNet built-in score in the observed network is shown as a contextual reference, together with the mean of our \DIshort{} implementation across ten randomized networks. The quantitative comparison reported in the text uses our implementation on both the observed and randomized networks, rather than comparing the levels of these two plotted series.}
    \label{fig:null-reflen-D}
\end{figure}

The randomized network reproduces a large part of the decline in \DIshort{} with reference count without reproducing the observed sign change.
Figure~\ref{fig:null-reflen-D} reports the decline using built-in \DIshort{} scores on the real citation network and our implementation of Equation~\eqref{eq:D} on the randomized citation networks. \DIshort{} declines by approximately 0.0165 from the first reference-count bin to its minimum in the observed network and by approximately 0.0141 in the randomized networks.
Thus about 85\% of the observed decline survives randomization.
In the randomized networks, the $\Nj$ share remains near zero because the co-citation structure that places later papers in $\Nj$ has been destroyed.
The $\Ni$ share nevertheless declines, so \DIshort{} approaches zero from above in the randomized network without crossing it.
This result does not establish that observed \DIshort{} contains no substantive signal.
It establishes the narrower point that a pronounced decline with reference count does not require differences in substantive disruption to arise.

\section{Reference-balancing algorithm and diagnostics}
\label{si:reference-balancing}

\begin{figure}[h]
    \centering
    \includegraphics[width=0.85\textwidth]{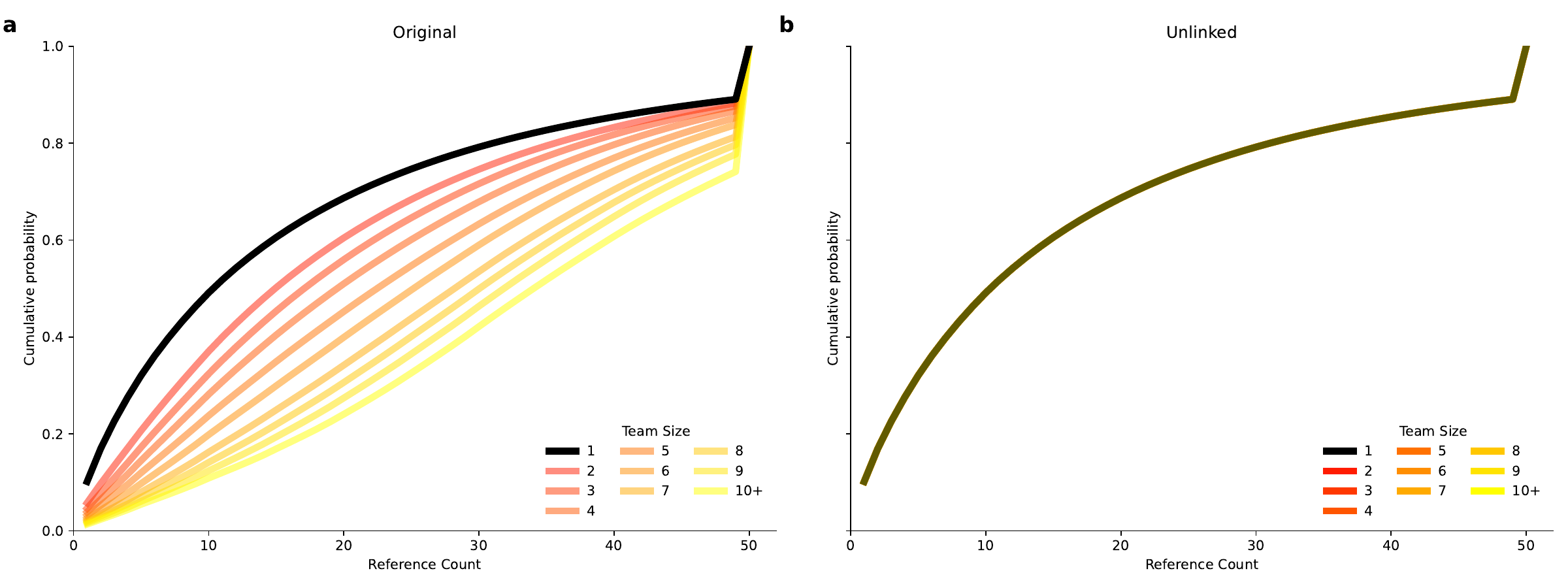}
    \caption{\textbf{Cumulative distributions of reference count before balancing.} 
    (a) Empirical cumulative distribution in the real database. At every nonterminal bin boundary used by the algorithm, the single-author CDF lies above the CDF of each larger team-size group. Counts of 50 or more are pooled in the terminal bin. This ordering makes the single-author distribution attainable through deletion alone at the binned resolution.
    (b) After reference balancing, the deletion procedure aligns the reference-count distribution of every larger team-size group with the single-author target. All curves collapse.
    }
    \label{fig:reflen_cdf_by_teamsize}
\end{figure}

The \nameref{sec:data-methods} Methods section describes the deletion-only reference-balancing design and its empirical stochastic-ordering condition.
Figure~\ref{fig:reflen_cdf_by_teamsize}a displays that condition. The remaining details below document its implementation and the achieved balance.

For each team-size group $g>1$, we construct a monotone probability-mass coupling between the group's observed distribution and the single-author target.
The coupling assigns mass only to transitions from an original count bin $a$ to a retained count bin $b\leq a$.
For a paper in group $g$ and original bin $a$, we draw a retained-reference bin from the coupling's conditional transition probabilities.
Below 50 references, the selected bin gives the exact number of references to retain.
Papers remaining within the pooled $50+$ tail are not thinned within that tail because the binned coupling does not identify their exact target counts.
The retained count is also bounded by the number of linked references available in the citation-network representation.
Conditional on that count, references are selected uniformly without replacement.
Because references are stored across distributed shards, the retained total is allocated among shards by sequential hypergeometric draws. This is equivalent to drawing one uniform subset from the paper's complete linked reference list and prevents storage location from affecting selection.
At the binned construction-population level, the transition matrix maps each source distribution exactly to the single-author target in expectation (Figure~\ref{fig:reflen_cdf_by_teamsize}b).

Absolute score levels are not directly comparable because deleting references changes the \DIshort{} calculation mechanically.

The supplementary figures below show nested samples extending through later decades for the shorter observable windows; these are cumulative samples, not independent cohorts.

\begin{figure}[h]
    \centering
    \includegraphics[width=0.7\textwidth]{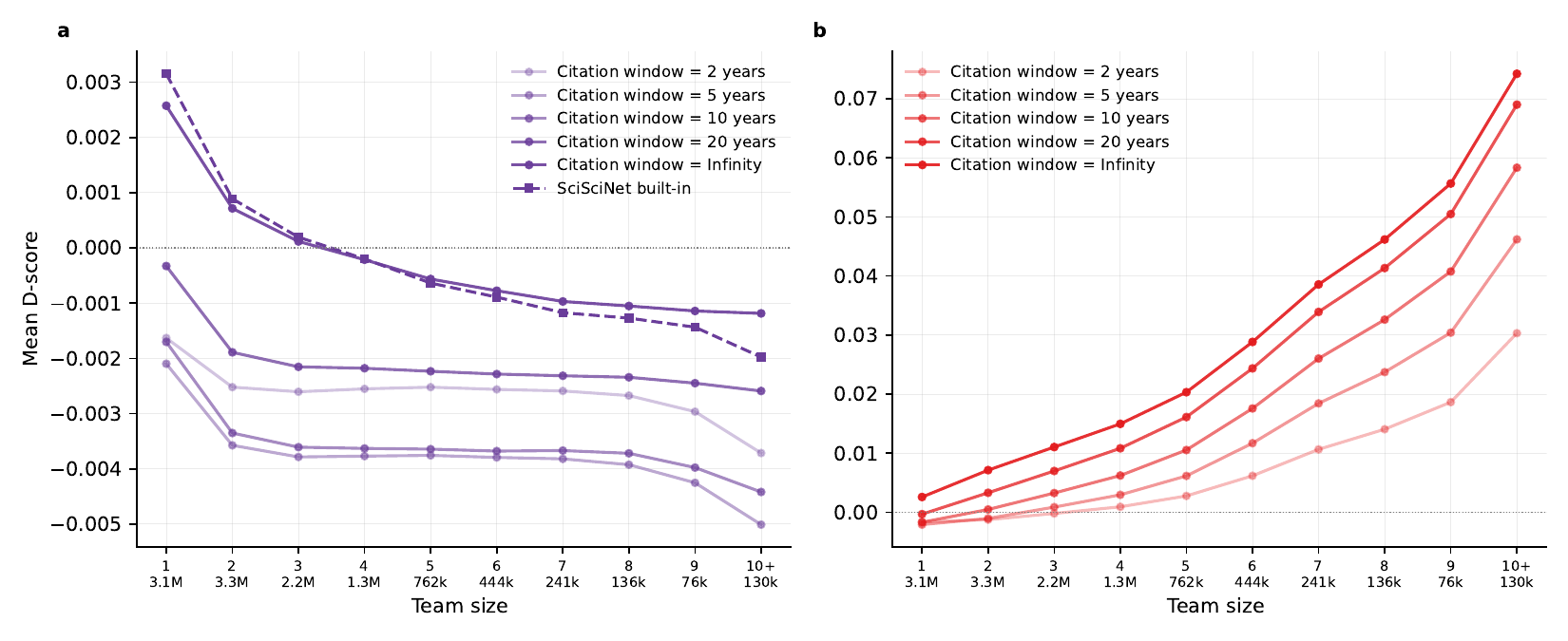}
    \includegraphics[width=0.7\textwidth]{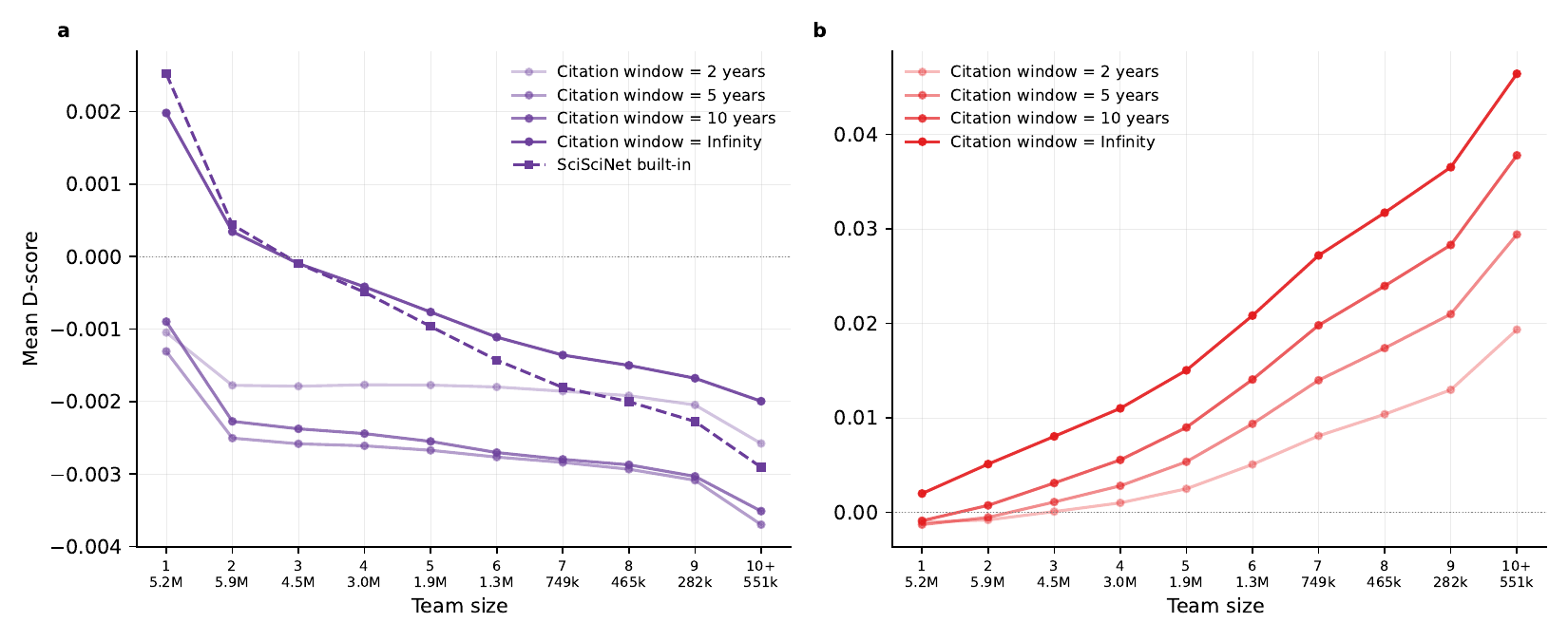}
    \includegraphics[width=0.7\textwidth]{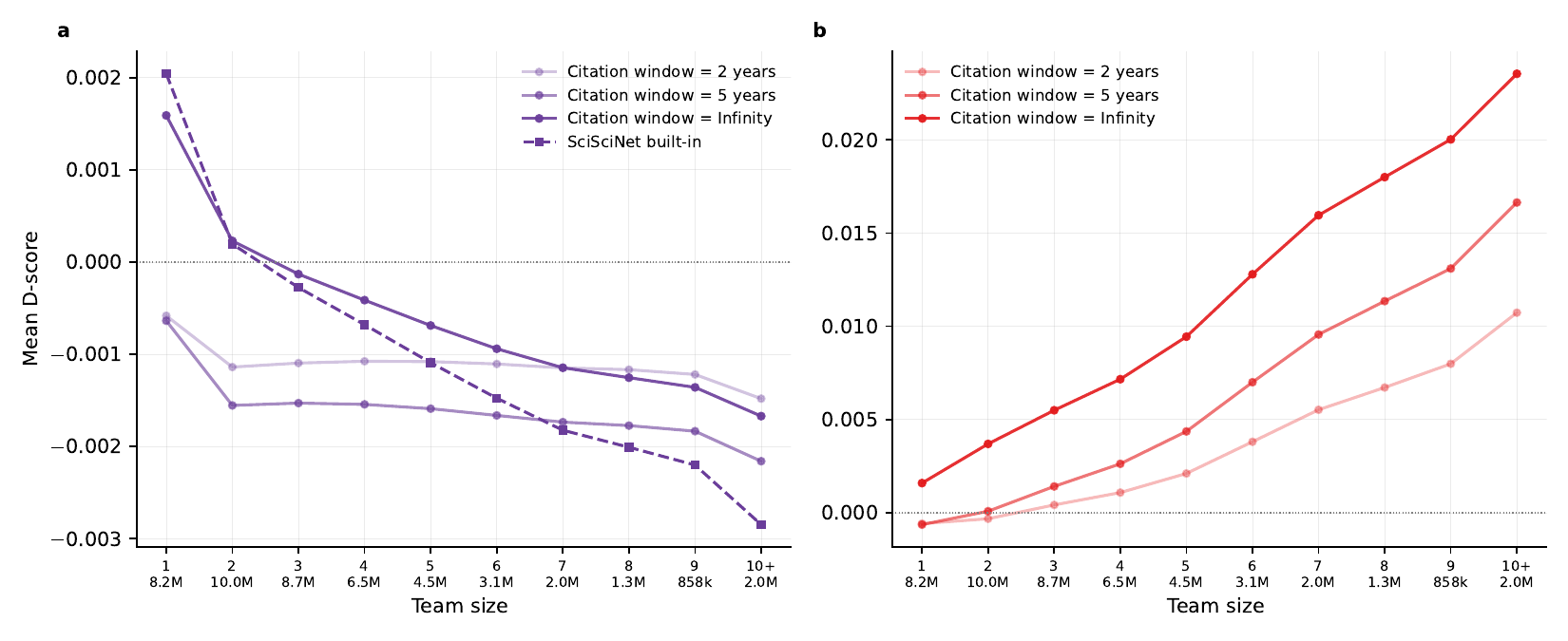}
    \caption{Mean \DIshort{} by team size before and after reference balancing, using a 20-year (top) , 10-year (mid), or 5-year (bottom) citation window and papers observable through the stated cohort.}
    \label{fig:debiased_result_window}
\end{figure}

\section{Exact \OneD{} and Monte Carlo convergence}
\label{si:oneD-convergence}

For focal paper $F$, let $X_F(r)=1D_F^{(r)}$ be the pairwise score associated with reference $r\in R_F$.
The exact estimand and the estimator used in the analyses are
\begin{align}
  \mu_F &= \frac{1}{|R_F|}\sum_{r\in R_F}X_F(r), \\
  \widehat\mu_{F,m} &= \frac{1}{m}\sum_{b=1}^{m}X_F(r_b),
\end{align}
where $r_1,\ldots,r_m$ are sampled independently and uniformly with replacement from $R_F$.
Conditional on the observed citation network and reference list,
\begin{equation}
  \mathbb{E}[\widehat\mu_{F,m}]=\mu_F,
  \qquad
  \operatorname{Var}(\widehat\mu_{F,m})=\frac{\sigma_F^2}{m},
\end{equation}
where $\sigma_F^2$ is the variance of $X_F(r)$ under uniform sampling from the paper's references.
The estimator is therefore unbiased and converges at the standard $m^{-1/2}$ rate.
Because every pairwise score lies in $[-1,1]$, its standard error is bounded above by $m^{-1/2}$ without distributional assumptions; empirical errors are considerably smaller.

Table~\ref{tab:auc-estimand} examines the aggregation order specified in the main text by comparing the AUC of the paper-level $m=30$ mean with the average of 30 single-draw AUCs.
The latter is not the estimand used in this study and is consistently lower.

\begin{table}[htbp]
\centering
\caption{Why validation uses the paper-level mean before calculating AUC. ``Paper mean'' calculates one AUC from each paper's mean across 30 draws. ``Mean single-draw AUC'' calculates an AUC for each draw and averages the 30 AUCs.}
\label{tab:auc-estimand}
\begin{tabular}{lrrr}
\toprule
Benchmark & Paper-mean AUC & Mean single-draw AUC & Difference \\
\midrule
Nobel & 0.9813 & 0.9045 & 0.0768 \\
\textit{Physical Review} Milestones & 0.9754 & 0.8802 & 0.0952 \\
\textit{Science} Breakthroughs & 0.9506 & 0.8419 & 0.1087 \\
Turing Award & 0.9425 & 0.8641 & 0.0784 \\
Lasker Award & 0.9785 & 0.8897 & 0.0888 \\
Test of Time & 0.9388 & 0.8248 & 0.1140 \\
\bottomrule
\end{tabular}
\end{table}

\begin{figure}[h]
  \centering
  \includegraphics[width=0.85\textwidth]{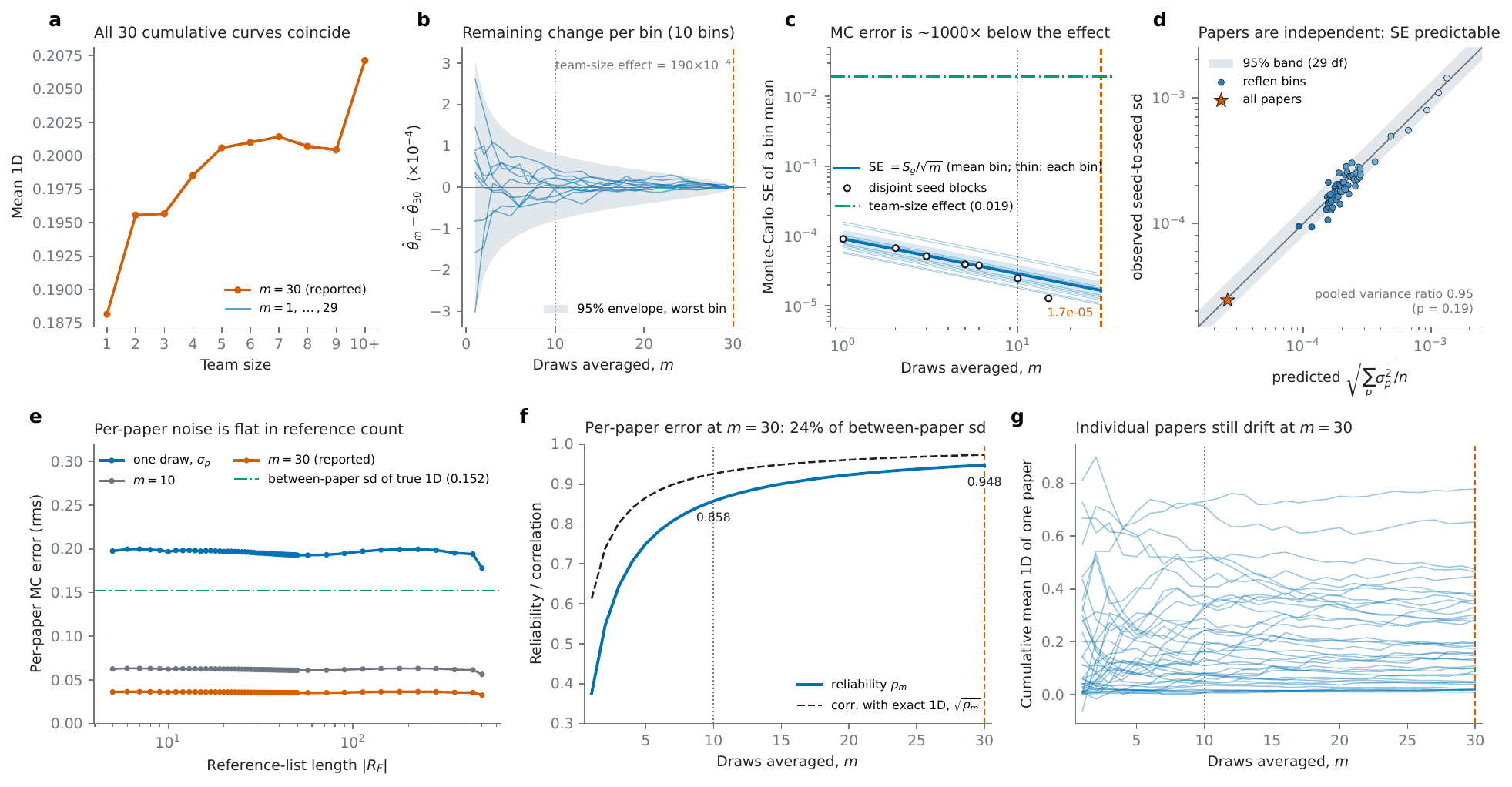}
  \caption{\textbf{Monte Carlo convergence of \OneD{}.} Thirty independent draws for each of focal papers with at least five references and at least one citation.
  (a)~Mean \OneD{} by team size after averaging the first $m$ draws, $m=1,\ldots,30$. The 30 curves coincide.
  (b)~Difference between the $m$-draw and 30-draw curves for each team-size group. The grey band is the 95\% envelope $1.96\,S_g\sqrt{1/m-1/30}$ for the noisiest group, where $S_g$ is the between-draw standard deviation of a group mean. For scale, the rise from team size
  1 to 10+ is $190\times10^{-4}$.
  (c)~Monte Carlo standard error of a team-size group mean, $S_g/\sqrt{m}$ (thin lines: individual groups; thick line: average; band: 95\% interval for $S_g$), with direct estimates from disjoint blocks of draws (circles). At $m=30$ the average error is
  $1.7\times10^{-5}$, about three orders of magnitude below the team-size effect (dash-dotted line).
  (d)~Observed between-draw standard deviation of group means against the value predicted from per-paper variances, $\sqrt{\sum_F\sigma_F^2}/n$, for reference-count groups and for all papers. The agreement (pooled variance ratio 0.95, $p=0.19$) indicates that draws are
  independent across papers, so the error of a mean over any $n$ papers is approximately $0.196/\sqrt{mn}$.
  (e)~Root-mean-square $\sigma_F$ by reference count for a single draw and after $m=10$ and $m=30$, against the between-paper standard deviation of $\mu_F$ (0.152).
  (f)~Reliability $\rho_m=\operatorname{Var}(\mu_F)/[\operatorname{Var}(\mu_F)+\mathbb{E}(\sigma_F^2)/m]$ of a paper's $m$-draw estimate, and its correlation $\sqrt{\rho_m}$ with $\mu_F$; $\rho_{30}=0.948$.
  (g)~Running averages for 40 randomly sampled papers.
  Dotted and dashed vertical lines mark $m=10$ and $m=30$.}
  \label{fig:si-mc-convergence}
  \end{figure}

Figure~\ref{fig:si-mc-convergence} reports diagnostics of Monte Carlo behavior.

\section{Reference count, citation windows, and \OneD{} robustness}
\label{si:oneD-robustness}
\label{si:oneD-window}

The one-reference construction removes a cumulative mechanism in the \DIfull{}: adding another reference does not automatically add that reference's citing papers to the same paper-level denominator.
It does not constrain the empirical distribution of focal-paper--reference relationships.
Mean \OneD{} therefore need not be flat as reference count changes, and in the observed data it increases across much of the reference-count range.
This association may reflect differences in the kinds of relationships contained in longer bibliographies.
Accordingly, we describe \OneD{} as removing the mechanical cumulative penalty, not as statistically independent of reference count.

\begin{landscape}
    \begin{figure}[p] %
        \centering
        \includegraphics[width=\linewidth]{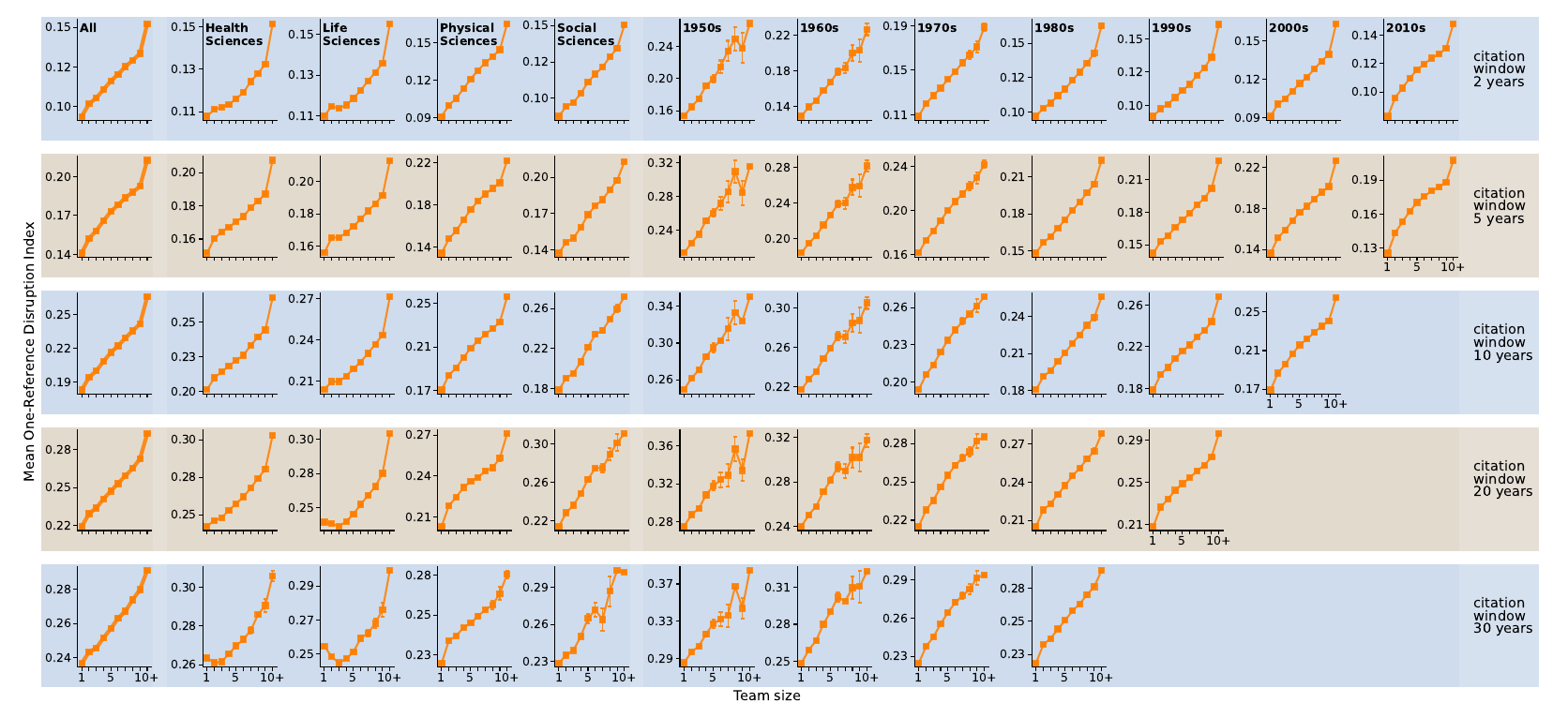}
        \caption{\textbf{Mean \OneD{} by team size using various citation windows}. Each row demonstrates a citation window, shown for the pooled sample, the four major scientific domains, and available publication decades.}
        \label{fig:oneD_teamsize_windows}
    \end{figure}
\end{landscape}

Figure~\ref{fig:oneD_teamsize_windows} report the finite-window robustness analyses described in~\nameref{sec:data-methods}.
Longer windows require complete follow-up and therefore restrict analysis to older focal-paper cohorts.
Each supplementary window figure therefore states its publication-year ceiling, and comparisons across windows must distinguish changes in follow-up from changes in cohort composition.

\section{Benchmark construction, attrition, and paired validation}
\label{si:validation}
\label{si:refcount-baseline}

The benchmark collections listed in \nameref{sec:data-methods} were assembled from published source lists, award records, and linked bibliographic information~\citep{2018-li,ioannidis2020work,2025-yy,2023-WangShiyun}. We constructed the Lasker collection from the key-publication lists on official Lasker Foundation pages for the Basic and Clinical Medical Research awards. 
We constructed the Test of Time collection from official retrospective paper-award lists covering 12 computer science conference series: CVPR, ICCV, ECCV, SIGMOD, PODS, VLDB, SIGCOMM, SIGIR, KDD, WWW, ICSE, and SOSP. The collection includes venue-specific Test of Time, Most Influential Paper, long-term impact, and Hall of Fame awards.
Records were deduplicated and mapped to OpenAlex works using DOI, title, publication year, and other available metadata.
Ambiguous matches were resolved using record-level provenance and adjudication.

Table~\ref{si:tab:benchmarks} reports source-list size and final eligible positives.
The eligible counts overlap across collections. Their sum is 1,469, while the unique union contains 1,399 papers.
The full common eligible population contains 41,152,454 papers.

\begin{table}[ht]
    \centering
    \caption{\textbf{Expert-curated breakthrough benchmarks in the paired validation.}}
    \label{si:tab:benchmarks}
    \begin{tabular}{lllllll}
    \toprule
    Benchmark  & \shortstack[l]{Number\\of papers} & \shortstack[l]{Number of \\ papers in\\ OpenAlex} & \shortstack[l]{Number of \\eligible \\ papers} & \DIshort{} AUC & \OneD{} AUC & \shortstack[l]{Reference \\ count AUC}\\
    \midrule
    \shortstack[l]{ACM A.M.\\ Turing Award}         & 79 & 54 & 38  & 0.792  & \textbf{0.942} & 0.311 \\
    \midrule
    Lasker Awards & 114 & 109 & 99  & 0.597     & \textbf{0.979} & 0.598 \\
    \midrule
    Nobel Prize  & 778 & 574 & 468 & 0.731     & \textbf{0.981} & 0.395 \\
    \midrule
    \shortstack[l]{\textit{Physical Review}\\ Milestone} & 300 & 291 & 270 & 0.398     & \textbf{0.975} & 0.522 \\
    \midrule
    \shortstack[l]{\textit{Science} \\Breakthroughs \\of the Year} & 334 & 325 & 318 & 0.259     & \textbf{0.951} & 0.650 \\
    \midrule
    Test of Time    & 334 & 319 & 276 & 0.553   & \textbf{0.939} & 0.519 \\   
    \bottomrule
    Total & 1,939 & 1,672 & 1,469 & 0.555 & \textbf{0.961} & 0.499 \\
    \bottomrule
    \end{tabular}
\end{table}

Of the 470 source records that are not eligible, the largest primary reason is absence from OpenAlex citation network (267 records), which is required to calculate either score, so this attrition reflects network coverage rather than a benchmark-specific exclusion.
The remaining primary reasons are fewer than five linked references (140), no SciSciNet built-in score (56), publication after 2019 (6), and no recorded citation (1).
Each record was assigned one primary reason according to a prespecified hierarchy. Because eligibility failures overlap, these counts are not marginal exclusions.
Because only $38$ Turing Award papers remain eligible, its AUC should be interpreted cautiously.

The paired analysis requires both built-in \DIshort{} and \OneD{}, at least five references, at least one citation, and publication before 2020.
The criteria are overlapping rather than a sequential partition: 15,667,776 papers lack a built-in score, 13,717,538 have fewer than five linked references, 21,376,577 have no citation, and 21,484,938 were published after 2019.
No eligible paper lacks \OneD{}.
Their intersection is the 41,152,454 paper common population used for all paired comparisons.

\paragraph{Reference-count diagnostic.}
Table~\ref{si:tab:benchmarks} reports the paired validation results and includes reference count as a diagnostic baseline.
Reference count is not fitted to the labels and does not discriminate consistently: its AUC is below 0.5 for Nobel Prize and ACM A.M. Turing Award papers, close to 0.5 for the \textit{Physical Review} Milestone and Test of Time collections, and above 0.5 for Lasker Awards and \textit{Science} Breakthroughs of the Year papers.
The simple average is approximately 0.50, but the heterogeneity across collections is more informative than that average.

Any strictly increasing transformation of reference count preserves these AUCs, while a strictly decreasing transformation replaces each AUC by approximately one minus its value.
Neither direction can produce consistently high discrimination across all six collections.
The strong and consistent \OneD{} results therefore cannot be reproduced by simply relabeling reference count as a disruption score.
This diagnostic addresses the possibility that \OneD{} merely recodes reference count. It does not establish that the six collections uniquely define disruption.
The collections identify work recognized by scientific communities, and such recognition can reflect influence, importance, or award conventions in addition to disruption.

\paragraph{Implementation concordance.}
On the common validation population, the AUC of our unrestricted \DIshort{} implementation differs from that of built-in \DIshort{} by no more than 0.030 across the six benchmarks.
This concordance supports its use for mechanism analyses while retaining the independently produced built-in score as the main outcome.